\documentclass[structabstract]{aa}
\pdfoutput=1
\usepackage{subfig,graphicx}
\usepackage{txfonts}
\usepackage{natbib}
\begin{document}

\title{GOODS-Herschel: dust attenuation properties of UV selected high redshift galaxies}

\author{V. Buat\inst{1}
   \and S. Noll\inst{2}
   \and D. Burgarella \inst{1}
   \and E. Giovannoli \inst{1,3}
   \and V. Charmandaris\inst{4,5,6}
   \and M. Pannella\inst{7}
   \and H.S. Hwang \inst{8}
   \and D. Elbaz \inst{7}
   \and M. Dickinson\inst{9}
   \and G. Magdis\inst{10}
   \and N. Reddy\inst{11}
   \and E.J. Murphy\inst{12}
        }

\offprints{V. Buat}

\institute{Aix-Marseille  Universit\'e,  CNRS, LAM (Laboratoire d'Astrophysique de Marseille) UMR7326,  13388, France\\ \email{veronique.buat@oamp.fr}
      \and Institut f\"ur Astro- und Teilchenphysik, Universit\"at Innsbruck, Technikerstr. 25/8, 6020 Innsbruck, Austria
      \and University of the Western Cape, Private Bag X17, 7535, Bellville, Cape Town, South Africa
      \and University of Crete, Department of Physics and Institute of Theoretical \& Computational Physics, GR-71003 Heraklion, Greece
      \and IESL/Foundation for Research \& Technology-Hellas, GR-71110 Heraklion, Greece
      \and Chercheur Associ\'e, Observatoire de Paris, LERMA (CNRS:UMR8112), 61 Av. de l'Observatoire,
F-75014, Paris, France
      \and Laboratoire AIM-Paris-Saclay, CEA/DSM/Irfu - CNRS - Universit\'e Paris Diderot, CE-Saclay, F-91191 Gif-sur-Yvette, France
      \and Smithsonian Astrophysical Observatory, 60 Garden St., Cambridge, MA 02138, USA
      \and National Optical Astronomy Observatory, 950 North Cherry Avenue, Tucson, AZ 85719, USA
      \and Department of Physics, University of Oxford, Keble Road, Oxford OX1 3RH, UK
      \and Department of Physics and Astronomy, University of California, Riverside, 900 University Avenue, Riverside, California 92521
      \and Spitzer Science Center, California Institute of Technology, Pasadena, CA, 91125, USA
           }

\date{Received ; accepted}


\abstract
   {Dust attenuation in galaxies is poorly known, especially at high redshift. And yet the amount of dust attenuation is a key parameter to deduce accurate star formation rates from ultraviolet (UV) rest-frame measurements. The wavelength dependence of the dust attenuation is also of fundamental importance to interpret the observed spectral energy distributions (SED) and to derive photometric redshifts or physical properties of galaxies.}
   {We want to study dust attenuation at UV wavelengths at high redshift, where the UV is redshifted to the observed visible light wavelength range. In particular, we search for a UV bump and related implications for dust attenuation determinations.}
   {We use photometric data in the $Chandra$ Deep Field South (CDFS), obtained in intermediate and broad band filters by the MUSYC project, to sample the UV rest-frame of 751 galaxies with $0.95 < z < 2.2$. When available, infrared (IR) $Herschel$/PACS\thanks{$Herschel$ is an ESA space observatory with science instruments provided by European-led Principal Investigator consortia and with important participation from NASA.} data from the GOODS-$Herschel$ project, coupled with {\it Spitzer}/MIPS measurements, are used to estimate the dust emission and to constrain dust attenuation. The SED of each source is fit using the CIGALE code. The amount of dust attenuation and the characteristics of the dust attenuation curve are obtained as outputs of the SED fitting process, together with other physical parameters linked to the star formation history.}
   {The global amount of dust attenuation at UV wavelengths is found to increase with stellar mass and to decrease as UV luminosity increases. A UV bump at 2175\,$\AA$ is securely detected in 20\% of the galaxies, and the mean amplitude of the bump for the sample is similar to that observed in the extinction curve of the LMC supershell region. This amplitude is found to be lower in galaxies with very high specific star formation rates, and 90$\%$ of the galaxies exhibiting a secure bump are at $z < 1.5$. The attenuation curve is confirmed to be steeper than that of local starburst galaxies for 20$\%$ of the galaxies. The large dispersion found for these two parameters describing the attenuation law is likely to reflect a wide diversity of attenuation laws among galaxies. The relations between dust attenuation, IR-to-UV flux ratio, and the slope of the UV continuum are derived for the mean attenuation curve found for our sample. Deviations from the average trends are found to correlate with the age of the young stellar population and the shape of the attenuation curve.}


\authorrunning {V. Buat et al.}
\titlerunning{Dust attenuation in high redshift galaxies}
\maketitle

%

\section{Introduction}\label{sec:intro}

Although dust is a minor component in galaxies by mass, its effect on the observation of their stellar populations is striking. Dust captures a large fraction of the stellar emission, especially at short wavelengths. This process makes the direct observation of stellar populations from the UV to the near-IR, where they emit their light, insufficient to recover all the emitted photons. Reliable dust corrections are mandatory for measuring the star formation rate in the universe and its evolution with redshift from UV-optical surveys. When dust emission is measured, accurate star formation rates can be derived by combining IR and UV data, but these data are often not available, in particular for deep optical surveys \citep[e.g.,][]{ilbert10}. As a consequence, it is particularly important to study the dependence of dust attenuation on parameters such as the observed luminosity, the stellar mass, or the slope of the UV continuum, since it could be used to correct large samples for the effect of dust attenuation, at least in a statistical way.

Any modelling of stellar populations in galaxies must also include attenuation from interstellar dust. Solving the radiation transfer in model galaxies is the best way to build physical and self-consistent SEDs. These models calculate the effective obscuration and produce attenuation curves as outputs, which are generally very different from the extinction curves affecting the flux of each star \citep[e.g.,][]{witt00,pierini04,tuffs04,panuzzo07}. However, these sophisticated models rely on numerous free parameters and physical assumptions that are difficult to constrain from the integrated emission from entire galaxies and for very large numbers of objects. Simpler models have been specifically developed to analyse large samples of galaxies, introducing attenuation curves, recipes, and/or templates. The number of free parameters is relatively small. These codes are often developed to measure photometric redshifts and physical parameters such as the star formation rate (SFR) and the stellar mass ($M_{\rm star}$). With the availability of mid and far-IR data for large samples of galaxies, new codes are emerging that combine stellar and dust emission on the basis of the balance between the stellar luminosity absorbed by dust and the corresponding luminosity re-emitted in the IR \citep{dacunha08,noll09b}.

Attenuation laws are introduced in all these codes, except for those which include a full radiation transfer treatment. The most popular attenuation curve is that of \citet{calzetti00}, built for local starburst galaxies. Some specific recipes such as a time dependence of dust attenuation are sometimes introduced \citep{charlot00,panuzzo07}.

The attenuation law for local starbursts, based on spectroscopic data, does not exhibit a bump at 2175\,$\AA$ such as that observed in the extinction curves of the Milky Way (MW) or the Large Magellanic Cloud (LMC) \citep{fitz07,gordon03}. The presence of a UV bump in the attenuation curve of galaxies remains an open issue. In the nearby universe, the UV wavelength range has been investigated thanks to GALEX observations, but the results remain controversial. \citet{wijesinghe11} analyse the consistency of SFR indicators based on GALEX measurements in the far-ultraviolet (FUV) and near-ultraviolet (NUV) bands and fluxes in the H$_\alpha$ line, and conclude that they must consider an obscuration curve without any 2175\,$\AA$ feature. Instead, from a careful analysis of pairs of galaxy SEDs, \citet{wild11} conclude that the UV slope of the attenuation curve is consistent with the presence of a bump at 2175\,$\AA$, a conclusion also reached by \citet{conroy10b} from an analysis of the GALEX-SDSS colours of galaxies.
At higher redshifts, several authors introduce a bump with moderate amplitude to improve photometric redshifts \citep{ilbert09,kriek11}. Direct evidence of bumps comes from the analysis of the galaxy spectra at $1 < z < 2.5$: Noll and collaborators analyse high quality spectra of $\sim$~200 galaxies and find significant bumps in at least 30$\%$ of the sources \citep{noll05,noll07,noll09a}. In an earlier paper, we analysed SEDs of 30 galaxies in a redshift range between 0.95 and 2.2, observed through intermediate band filters and with IR detections from $Herschel$/PACS, and found evidence for a UV bump in the dust attenuation curve of all these galaxies \citep[][hereafter paper~I]{buat11b}. Selecting high-redshift galaxies on their observed optical colours is quite common (e.g. Lyman break or BzK galaxies \citep{reddy08,daddi05}, but most of the time mid- and far-IR data are not available for individual targets, making it difficult to obtain any direct measure of dust attenuation. Studies of dust emission often rely on stacking analyses \citep{rigopoulou10,burgarella11,reddy12}. The sensitivity of $Herschel$ in the deepest fields allows us to combine stellar and dust emission for individual galaxies at high redshift \citep{wuyts11}, and to perform SED fitting accounting for both components (paper~I).

In this work, we will select galaxies at optical wavelengths with a very good sampling of their UV rest-frame SED: essentially, the sample is UV selected with a redshift range between 1 and 2. When available, IR emission from $Herschel$/PACS will be added to better constrain dust attenuation. SED fitting will be performed to estimate the main characteristics of dust attenuation. The sample and the SED fitting process are described in Sects.~\ref{sec:sample} and \ref{sec:SED-fitting}, respectively. The global amount of dust attenuation and its variation with stellar mass and UV luminosity are discussed in Sect.~\ref{sec:attenuation}. Section~\ref{sec:attcurve} is devoted to the description of the attenuation law. In Sect.~\ref{sec:att_slope}, we revisit the relation between dust attenuation and the slope of the UV continuum. Our conclusions are presented in Sect.~\ref{sec:conclusions}. All magnitudes are given in the AB system. We assume that $\Omega_{\rm m} = 0.3$, $\Omega_{\Lambda} = 0.7$, and $H_0 = 70\,{\rm km\,s^{-1}\,Mpc^{-1}}$.

\section{Galaxy sample and rest-frame UV continuum}\label{sec:sample}

\subsection{Multi-wavelength data}\label{sec:multiwave}

As part of the GOODS-$Herschel$ key programme \citep{elbaz11}, the $Herschel$ Space Observatory \citep{pilbratt10} surveyed part of the Great Observatories Origins Deep Survey Southern field (GOODS-S): $10' \times 10'$ centred on the CDFS were observed at 100 and 160\,$\mu$m over 264\,hours by the PACS instrument \citep{poglitsch10}. These are the deepest images of the universe ever obtained with PACS.

\citet[][the MUSYC project]{cardamone10} compiled a uniform catalogue of optical and IR photometry for sources in GOODS-S, incorporating the GOODS {\it Spitzer} IRAC and MIPS data \citep{dickinson03} and optical WFI broad-band data obtained at the ESO MPG 2.2\,m telescope \citep{hildebrandt06}. In addition, they used deep intermediate-band imaging from the Subaru telescope to provide photometry with finer wavelength sampling than is possible from standard broad-band data. This was done to enable more accurate photometric redshifts, and it provides a valuable means of tracing the detailed shape of the UV rest-frame spectrum. The GOODS-S field defined for the MUSYC project is larger than that observed by $Herschel$. In order to be able to put constraints on the emission of the sources in the far-IR, we restrict the field to that observed by the GOODS-$Herschel$ programme.

In paper~I, we selected 30 galaxies that are strongly detected from the UV to the far-IR, and whose rest-frame UV was sampled by several intermediate-band filters. In the present work, we select a much larger sample of galaxies by relaxing the constraint on the wavelength coverage and keeping galaxies without IR data. We start with the MUSYC catalogue, selecting sources with a spectroscopic redshift between 0.95 and 2.2. In this redshift range we have more than ten photometric bands available in the UV rest-frame and a good sampling around 2175\,$\AA$. As in paper I, we consider all the optical broad bands ($U$, $U38$, $B$, $V$, $R$, $I$, $z$) and intermediate-band filters whose 5\,$\sigma$ depth was fainter than 25\,ABmag ($IA-427$, 445, 484, 505, 527, 550, 574, 598, 624, 651, 679, and 738 (\citet[][Table~2]{cardamone10}; paper~I, Fig.~1). We apply the same selection criteria as in paper~I to ensure that sources have data below rest-frame $1800\,\AA$: a 5\,$\sigma$ detection in the $U$ and $U38$ filters for sources at $z < 1.6$, and in the $B$ band and one intermediate filter corresponding to $\lambda < 1800\,\AA$ in the galaxy rest-frame for $z > 1.6$. We further restrict the sample to galaxies without a X-ray detection according to \citet{cardamone08} to minimise the contribution of an AGN to either the IR or UV emission of the sources. Our sample contains 751 sources, all detected by IRAC at 3.6\,$\mu$m. 726, 398, and 353 sources are detected at 4.5, 5.8, and 8\,$\mu$m, respectively.

We cross-correlate these 751 sources with the GOODS-$Herschel$ catalogue obtained from source extraction on the PACS images performed at the prior position of ${\it Spitzer}$ 24\,$\mu$m sources, as described in \citet{elbaz11}. IRAC coordinates are used for the Spitzer 24\,$\mu$m sources, and we adopt a tolerance radius between IRAC and optical coordinates equal to 1\,arcsec. 290 sources have a 3\,$\sigma$ detection at 24\,$\mu$m. We check that all these sources have a single MIPS counterpart within a radius of 3 arcsec. 76 sources are detected at both 24\,$\mu$m and 100\,$\mu$m, and 30 at 24\,$\mu$m, 100\,$\mu$m, and 160\,$\mu$m.

During our analysis, we will distinguish between galaxies that do and do not have detections of dust emission at 24\,$\mu$m (hereafter MIPS data) and 100\,$\mu$m (hereafter PACS data). Given the small number of sources with 160\,$\mu$m detections, we will not isolate the galaxies also detected at this wavelength from those detected at 100\,$\mu$m.

\begin{figure}
\centering
\includegraphics[width=9cm]{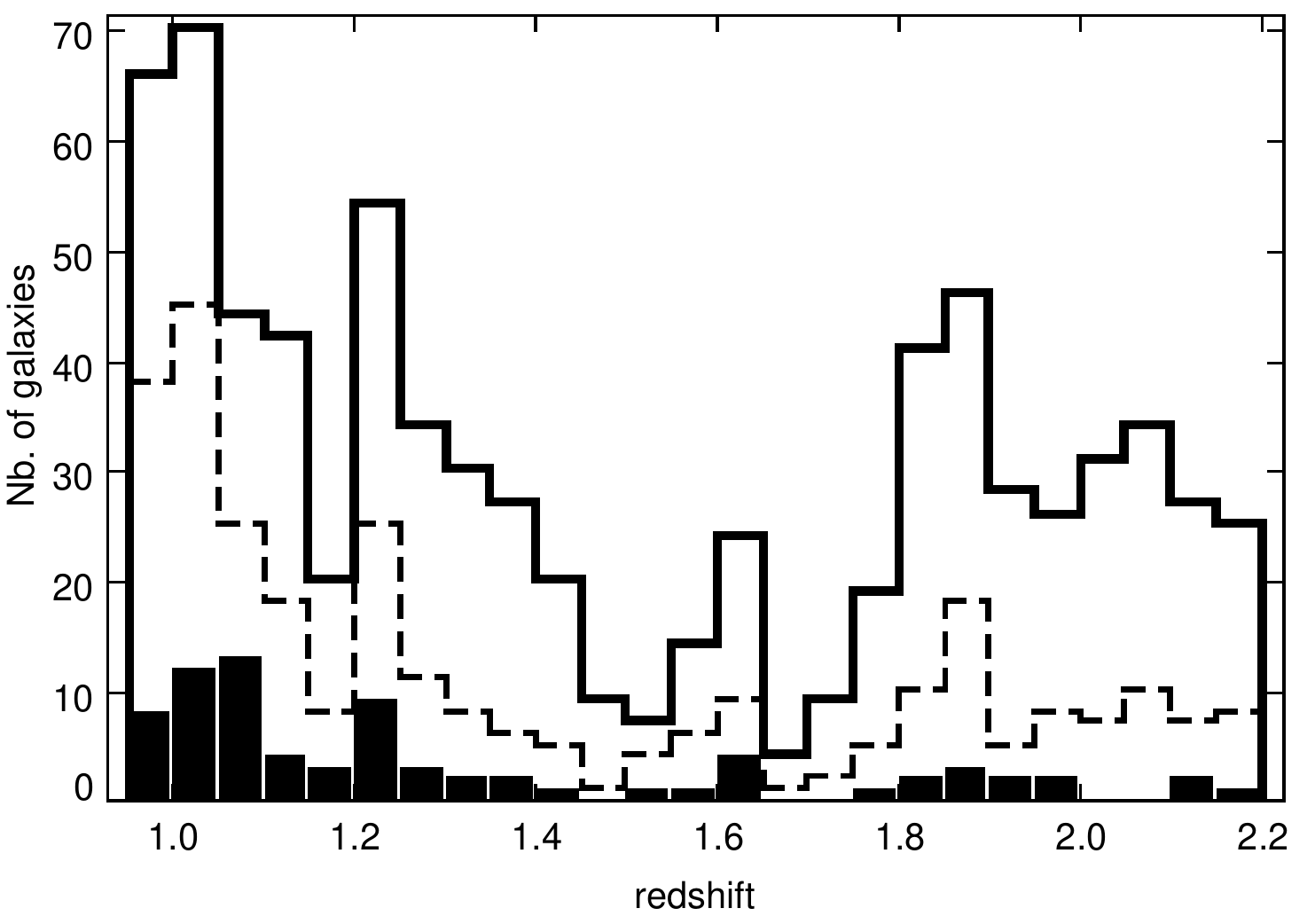}
\caption{Redshift distribution of the galaxy sample. The distribution for the whole sample is plotted as a solid line, that for the galaxies detected by MIPS with a dashed line, and the filled histogram corresponds to the redshifts of sources detected with PACS.}
\label{fig:zdist}
\end{figure}

The redshift distribution of the sources is shown in Fig.~\ref{fig:zdist}. The bimodality comes from the original distribution of spectroscopic redshifts collected by \citet{cardamone10} and mainly reflects the difficulty to measure spectroscopic redshifts for galaxies at $1.4 < z  < 1.8$. 55$\%$ of the galaxies selected have a redshift $z < 1.5$, and that percentage increases to 66 and 76$\%$ for the galaxies detected by MIPS and PACS respectively.

\begin{table*}
\caption{Input parameters for SED fitting with CIGALE}
\label{tab:parameters}
\centering
\begin{tabular}{c c c}
\hline
Parameter & Symbol & Range \\
\hline\hline
Dust component & & \\
\hline
distribution of dust mass & $\alpha$ & 1, 1.5, 2, 2.5\tablefootmark{1} \\
amplitude of the bump & $E_{\rm b}$ & 0, 0.5, 1, 1.5, 2.0, 2.5, 3, 3.5 \\
full width half maximum of the bump profile & $\gamma$ & 300, 350\,$\AA$\tablefootmark{2} \\
steepness of the attenuation law & $\delta$ & -0.8, -0.6, -0.4, -0.2, 0, 0.2 \\
$A_V$ for the young stellar population & $A_V ({\rm ySP})$ & 0.25, 0.5, 0.75, 0.9, 1.05, 1.2, 1.35, 1.5, 1.65, 1.8\,mag \\
reduction of $A_V$ for the old population &$f_{\rm att}$ & 0.25, 0.5, 1\tablefootmark{3} \\
\hline
Stellar component & \\
\hline
age of the old stellar population & $t_1$ & 1, 2, 3, 4, 5\,Gyr\tablefootmark{4} \\
$e$-folding rate of the old stellar population & $\tau_1$ & 1, 3\,Gyr \\
age of the young stellar population & $t_2$ & 0.01, 0.03, 0.1, 0.3\,Gyr \\
stellar mass fraction due to the young stellar population & $f_{\rm ySP}$ & 0.01, 0.02, 0.05, 0.1, 0.2, 0.5 \\
\hline
\end{tabular}
\tablefoot{Values of input parameters defining the dust emission and attenuation recipes (first part of the table) and the stellar populations (second part of the table). \\
\tablefootmark{1}{A single value of $\alpha$ ($\alpha = 2$) is used when PACS data are not available.} \\
\tablefootmark{2}{A single value of $\gamma$ is used for each run of the code. The baseline value is 350\,$\AA$.} \\
\tablefootmark{3}{A single value of $f_{\rm att}$ is assumed for each run of the code. The baseline value is 0.5.} \\
\tablefootmark{4}{The age of the old stellar population is always lower than that of the universe at the redshift of the object.}
}
\end{table*}

\subsection{UV continuum}\label{sec:UV}

The UV continuum of star-forming galaxies is commonly modelled by a power-law, $ f_\lambda ({\rm erg\,cm^{-2}\,s^{-1}\,\AA^{-1}}) \propto \lambda^{\beta}$, where $\beta$ is called the slope of the UV continuum. The observational determination of $\beta$ is made either from spectra or from broad-band photometry, sometimes with large uncertainties depending on the broad-band filters considered \citep{buat11b}. The wavelength range over which $\beta$ is measured also varies substantially \citep{calzetti01}. Because of the very good sampling of the UV rest-frame continuum of our sources, we can accurately measure the slope of the UV continuum from 1300 to 3000\,$\AA$ by performing a linear regression. The original wavelength range adopted by \citet{calzetti94} for fitting the UV continuum slope is $1200 - 2600$\,$\AA$. \citet{Meurer99} adopted a narrower range, $1200 - 1800$\,$\AA$. However, given our available photometric data, extending the range to 3000\,$\AA$ adds extra photometric points, which enable us to better constrain the UV continuum slope without significantly biasing the results that would be obtained from a shorter wavelength range. We start at 1300\,$\AA$ to avoid the flattening of the UV continuum at shorter wavelengths, down to 1200\,$\AA$ \citep{leitherer02,buat02}. $\beta$ is calculated in two ways, with and without data corresponding to the wavelength of the UV bump, and both estimates are found to be very similar. This is due to the fact that we have a large number of photometric data for each source with only a few measures inside the bump area. In the following, we will use the values of $\beta$ obtained with the full data set from 1300 to 3000\,$\AA$. The standard error on $\beta$ is typically found to be of the order of 0.2. The rest-frame luminosity at 1530\,$\AA$ of each galaxy is obtained by interpolation, using the result of the linear regression and the spectroscopic redshift of the source. In the following, we will define $L_{\rm FUV}$ as $\nu L_{\nu}$ at 1530\,$\AA$.

\section{SED fitting with CIGALE}\label{sec:SED-fitting}

\begin{figure}
\centering
\includegraphics[width=9cm]{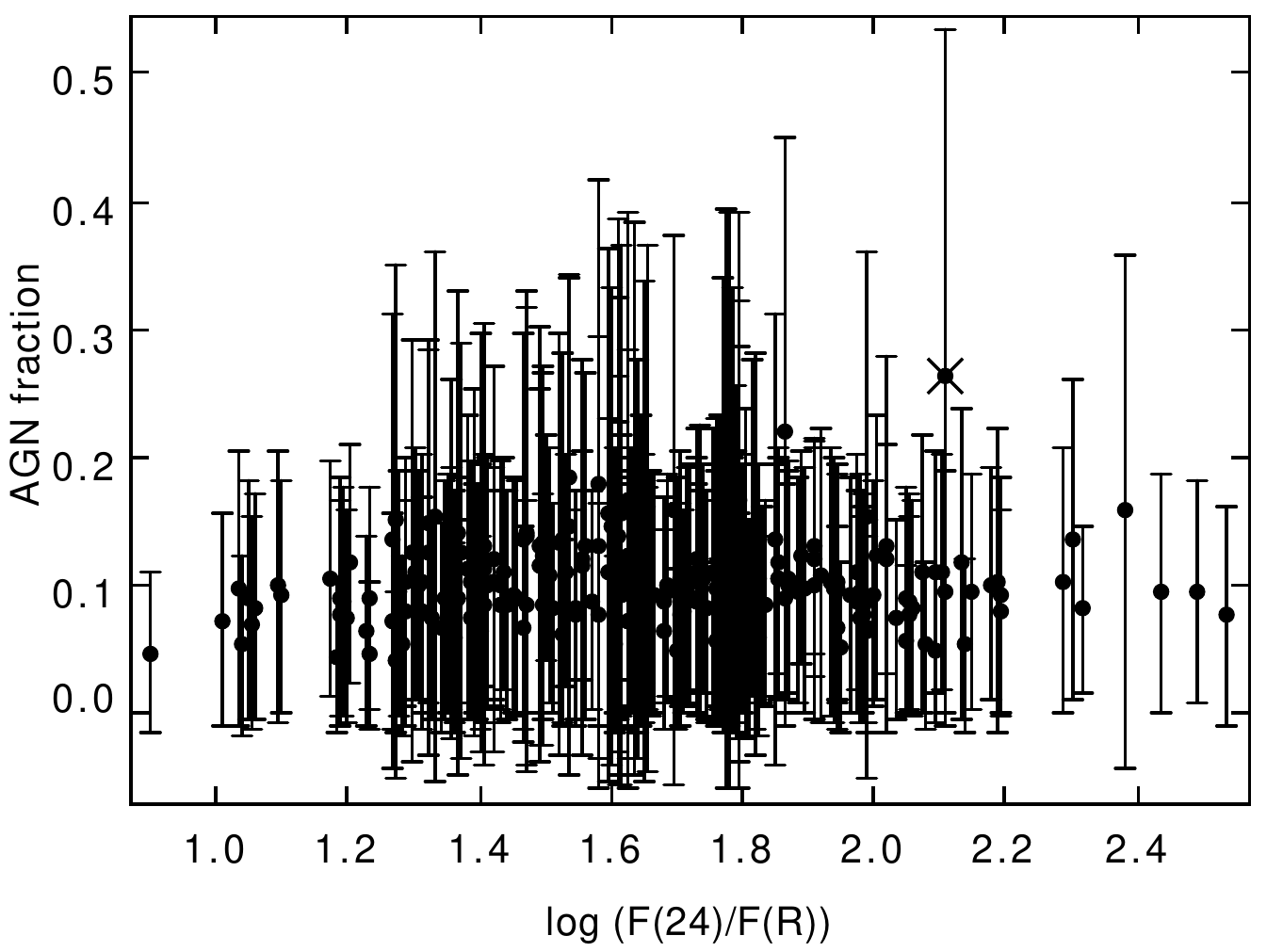}
\caption{Fraction of $L_{\rm IR}$ attributed to an obscured AGN by CIGALE as a function of the flux ratio at 24\,$\mu$m and in the $R$ band, the error bars represent the standard deviation calculated by CIGALE for this parameter. The cross indicates the only object with $F(24)/F(R) > 200$ and an AGN fraction higher than 0.15}
\label{fig:AGN}
\end{figure}

\begin{figure*}
\centering
\includegraphics[width=20cm]{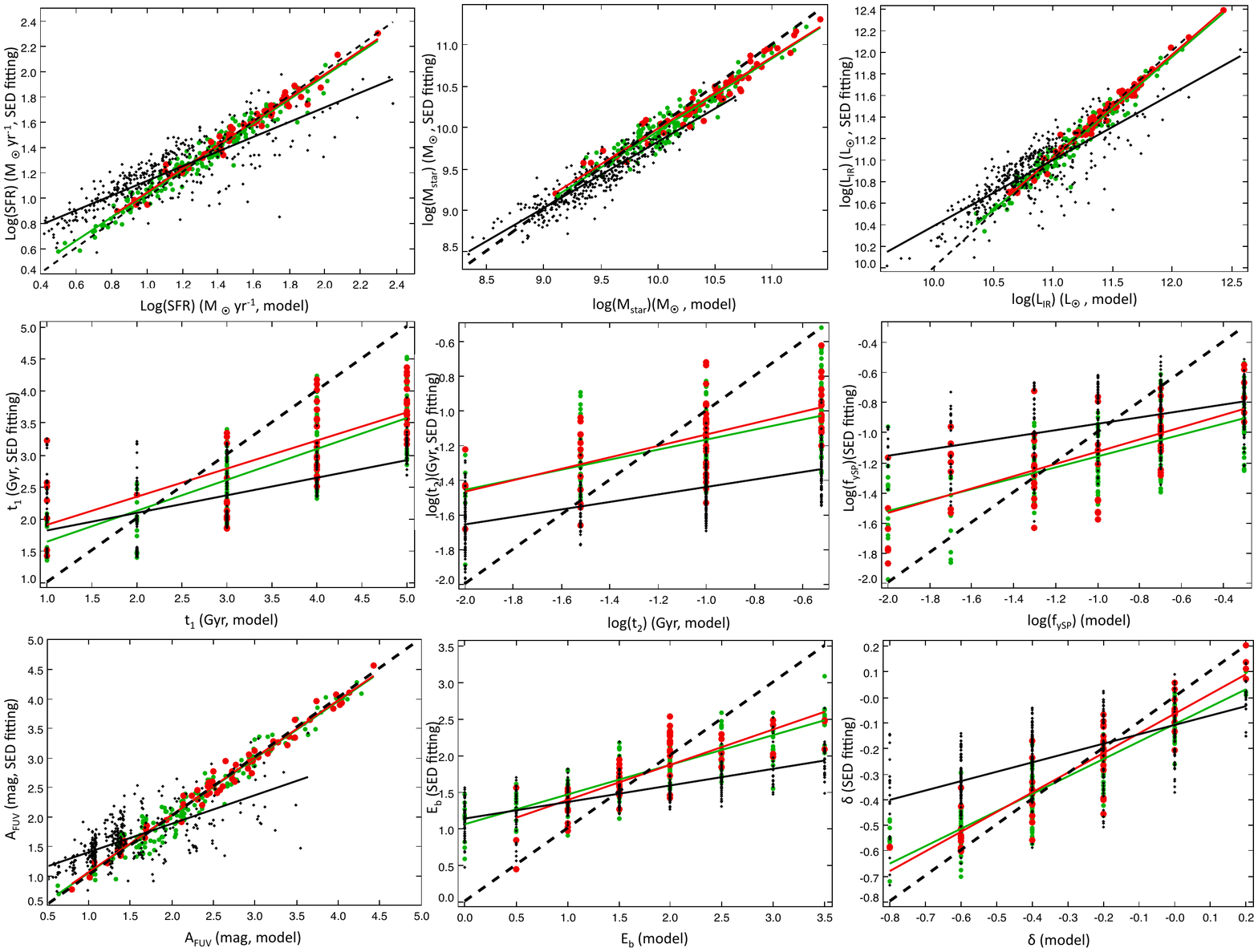}
\caption{Comparison of parameters estimated by SED fitting for the catalogue of artificial galaxies. The initial values of the parameters are on the x-axis and the values estimated after SED fitting on the y-axis. The galaxies without IR data are plotted as black dots, the galaxies with only a MIPS detection as green filled circles, and those also detected with PACS as large red filled circles. The regression lines for each subsample are also plotted as solid lines and the 1:1 relation as a dashed line. From the left top to the right bottom, the parameters considered are the SFR, $M_{\rm star}$, $L_{\rm IR}$, $t_1$, $\log(t_2)$, $\log(f_{\rm ySP})$, $A_{\rm FUV}$, $E_{\rm b}$, and $\delta$.}
\label{fig:mock}
\end{figure*}

\subsection{The CIGALE code}\label{sec:CIGALE}

SED fitting is performed with the CIGALE code (Code Investigating GALaxy Emission)\footnote{http://cigale.oamp.fr} developed by \citet{noll09b}. This is a physically-motivated code that derives properties of galaxies by fitting their UV-to-IR SEDs. CIGALE combines a UV-optical stellar SED with a dust component emitting in the IR and fully conserves the energy balance between the dust absorbed stellar emission and its re-emission in the IR. We refer to \citet{giovannoli11} for details on the use of CIGALE to fit SEDs of actively star-forming distant galaxies observed in their rest-frame UV and IR. In the present work, the sampling of the UV range is very good, so we expect to constrain the dust attenuation curve. To model the attenuation by dust, the code uses as a baseline the attenuation law of \citet{calzetti00}, and offers the possibility of varying the steepness of this law and adding a bump centred at 2175\,$\AA$. We refer to \citet{noll09b} for a complete description of the dust attenuation prescription. In brief, the dust attenuation is described as
\begin{equation}\label{eq:attlaw}
A(\lambda) = {A_V \over 4.05}~ (k'(\lambda) + D_{\lambda_0,\gamma,E_ {\rm b}}(\lambda)) \left(\lambda \over {\lambda_V} \right)^{\delta},
\end{equation}
where $\lambda_V = 5500\,\AA$, $k'(\lambda)$ comes from \citet{calzetti00} (Eq.~4) and $D_{\lambda_0,\gamma,E_{\rm b}}(\lambda)$, the Lorentzian-like Drude profile commonly used to describe the UV bump \citep{fitz90,noll09a}, is defined as
\begin{equation}\label{eq:bump}
D_{\lambda_0,\gamma,E_{\rm b}} = {{E_{\rm b} \lambda^2 \gamma^2} \over {(\lambda^2-\lambda_0^2)^2 + \lambda^2 \gamma^2}}.
\end{equation}
The factor $\left(\lambda \over {\lambda_V} \right)^{\delta}$ produces different slopes without modifying the visual attenuation $A_V$. It implies changes in the value of the effective total obscuration $R_V$, originally equal to 4.05 for the Calzetti et al. law \citep{noll09b,buat11b}.

The peak amplitude above the continuum, full width at half maximum, and central wavelength of the bump ($E_ {\rm b} $, $\gamma$, and $\lambda_0$) are free parameters of the code\footnote{$E_{\rm b}$ and $\gamma$ are related to the original parameters $c_3$ and $\gamma_{\rm FM}$ introduced by \citet{fitz90}: $E_{\rm b} = c_3/\gamma_{\rm FM}^2$ and $\gamma = \gamma_{\rm FM} \times \lambda_0^2$ }. We fix the central wavelength of the bump $\lambda_0$ at 2175\,$\AA$. The full width at half maximum of the feature $\gamma$ is not constrained by the SED fitting process. We adopt a value of 350\,$\AA$ intermediate between the value found for the MW and LMC and that found by \citet{noll09a} for high-redshift systems. We also ran CIGALE with $\gamma = 300\,\AA$ and checked that the exact value of this parameter has no influence on the determination of the other parameters. The input values for the other parameters are listed in Table~\ref{tab:parameters}. The range of values for each free parameter was chosen after many trials: we began with a very large range of values and then reduced the range by excluding values never chosen during the $\chi^2$ minimisation. We will re-discuss the choice of the dust attenuation parameters in Sects.~\ref{sec:attenuation} and \ref{sec:attcurve}.

We adopt the stellar populations synthesis models of \citet{maraston05}. We consider a stellar population experiencing an exponentially decreasing star formation, whose age and $e$-folding time are free parameters. A younger stellar population with a constant SFR is added to the primary population. We adopt a \citet{kroupa01} initial mass function (IMF). The two stellar components are linked by their mass fraction $f_{\rm ySP}$ and an attenuation correction is applied to each of them. The same attenuation law is assumed for both stellar populations and a reduction factor of the visual attenuation (expressed in magnitude), $f_{\rm att}$, is applied to the old stellar population to account for the distributions of stars of different ages \citep{charlot00, panuzzo07}. For our baseline model, $f_{\rm att}$ is fixed to 0.5, but we also performed runs with other values to check the validity of this assumption, since this parameter is ill-constrained (cf. Sect.~\ref{sec:attenuation}).

Dust luminosities (L$_{\rm IR}$ between 8 and 1000\,$\mu$m) are computed by fitting \citet{dale02} templates and are linked to the attenuated stellar population models (the stellar luminosity absorbed by the dust is re-emitted in the IR). The validity of \citet{dale02} templates for measuring total IR luminosities of sources detected by $Herschel$ is confirmed by the studies of \citet{elbaz10,elbaz11}. When PACS data are not available, as is the case for the majority of the sources, only one of the templates of \citet{dale02} is used, which corresponds to the exponent of the distribution of dust mass over heating intensity, $\alpha$, equal to 2. This is the mean value of $\alpha$ obtained for the subsample of galaxies with PACS detection. This template is very close to that used by \citet{wuyts08}. Using $\alpha = 2$ to extrapolate the predicted fluxes in the PACS bands gives values consistent with the PACS flux limits from \citet{elbaz11} for 88$\%$ of the sample at 100\,$\mu$m (flux limit: 0.8\,mJy) and the percentage increases to 92$\%$ at 160\,$\mu$m (flux limit: 2.4\,mJy). Adopting different values of $\alpha$ modifies these percentages. With $\alpha = 2.5$ corresponding to more quiescent and colder galaxies, only 6$\%$ of the predicted fluxes for galaxies not detected with PACS are found above the flux limits at 100 and 160\,$\mu$m. For a hotter dust emission with $\alpha = 1.75$, the fraction of galaxies which would have been detected by PACS reaches 18$\%$ at 100\,$\mu$m and 10$\%$ at 160\,$\mu$m. Therefore we do not favour low values of $\alpha$ for the galaxies not detected in IR.

We run CIGALE on the full sample of 751 galaxies. The input parameters as well as additional output parameters like the stellar mass, the instantaneous SFR, or the total IR luminosity are estimated from the probability distribution function (PDF). The expectation value and its standard deviation are derived for each parameter. 84$\%$ of the sample is fitted with a minimum value of the reduced $\chi^2$ lower than 3. The list of free parameters is given in Table~\ref{tab:parameters}.

 Although we discarded sources with an X-ray detection (Sect.~\ref{sec:multiwave}), highly obscured AGN could be present in our galaxy sample. To check this issue, we use the $F(24)/F(R)$ flux ratio: highly obscured Compton thick AGN are expected to exhibit $F(24)/F(R)$ flux ratios larger than few hundreds \citep{fiore09}. We also run CIGALE adding an AGN contribution as in paper~I (with two obscured AGN templates from the \citet{siebenmorgen04} library). In Fig.~\ref{fig:AGN}, we present the fraction of L$_{\rm IR}$ attributed to an obscured AGN by CIGALE as a function of $F(24)/F(R)$. The highest value of $F(24)/F(R)$ is 340. If we focus on galaxies with $F(24)/F(R) > 200$, only one source is found with an AGN fraction larger than 0.15. We discard this source from the following analysis. For the remaining sources, the average AGN fraction found by CIGALE is $0.09 \pm 0.07$. Since this low value is consistent with no AGN contribution, we will use the results of the SED fitting process without any AGN template in the following analysis.

\subsection{Analysis of a catalogue of artificial galaxies}\label{sec:mock}

In order to check the robustness and accuracy of the parameter estimation, we generate a catalogue of artificial SEDs from the input catalogue. First, we run the code on the full dataset to select the best model for each galaxy that corresponds to a single SED. The catalogue of artificial sources is built by integrating the best SED of each galaxy in the observed bands. A random error, typically of the order of 10$\%$, is added to the obtained fluxes. At the end, we have a sample of artificial galaxies whose properties and SEDs are expected to be representative of the real sources. The code is run on this artificial catalogue to compare the exact values of the physical parameters corresponding to the artificial SEDs to the parameters estimated by the code. With this catalogue, we want to check our ability to estimate individual output parameters of the code given the wavelength coverage. We do not attempt to measure the effects of different signal-to-noise ratios.

\begin{table}
\caption{Linear correlation coefficient ($R$) between the exact value of parameters and the value estimated by the code and its error (standard deviation $\sigma$ of the PDF) of the estimated values for the sample of artificial galaxies.}
\label{tab:correlation}
\centering
\begin{tabular}{l l l l l l l}
\hline
Parameters & \multicolumn{2}{c}{No IR data} & \multicolumn{2}{c} {MIPS data} & \multicolumn{2}{c} {PACS data} \\
& $R$ & $\sigma$ & $R$ & $\sigma$ & $R$ & $\sigma$ \\
\hline
$\log(SFR)~[M_{\odot}{\rm yr}^{-1}]$ & 0.86 & 0.25 & 0.99 & 0.11 & 0.99 & 0.09 \\
$\log(M_{\rm star})~[M_{\odot}]$ & 0.92 & 0.21 & 0.96 & 0.15 & 0.96 & 0.15 \\
$\log(L_{\rm IR})~[L_{\odot}]$ & 0.84 & 0.42 & 0.98 & 0.12 & 0.99 & 0.08 \\
\hline
$t_1$ [Gyr] & 0.67 & 0.98 & 0.77 & 1.02 & 0.77 & 1.00 \\
$\log(t_2)$ [Gyr] & 0.64 & 0.41 & 0.59 & 0.40 & 0.67 & 0.39 \\
$log(f_{\rm ySP})$ & 0.53 & 0.45 & 0.65 & 0.44 & 0.65 & 0.43 \\
\hline
$A_{\rm FUV}$ [mag] & 0.75 & 0.54 & 0.99 & 0.28 & 0.99 & 0.24 \\
$E_{\rm b}$ & 0.72 & 1.00 & 0.84 & 0.89 & 0.82 & 0.81 \\
$\delta$ & 0.68 & 0.24 & 0.90 & 0.17 & 0.93 & 0.14 \\
\hline
\end{tabular}
\end{table}

\begin{figure}
\centering
\includegraphics[width=8.2cm]{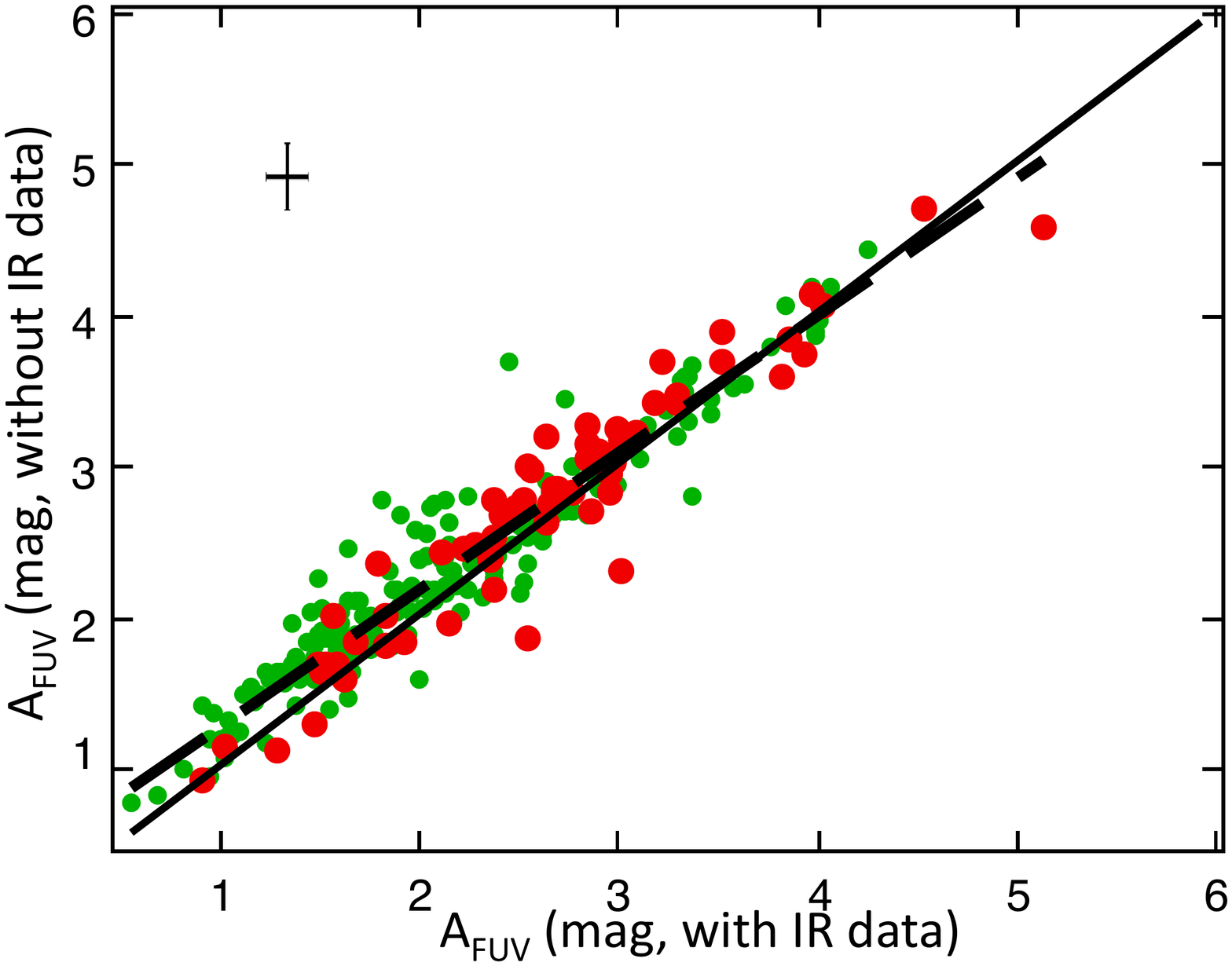}
\includegraphics[width=8.2cm]{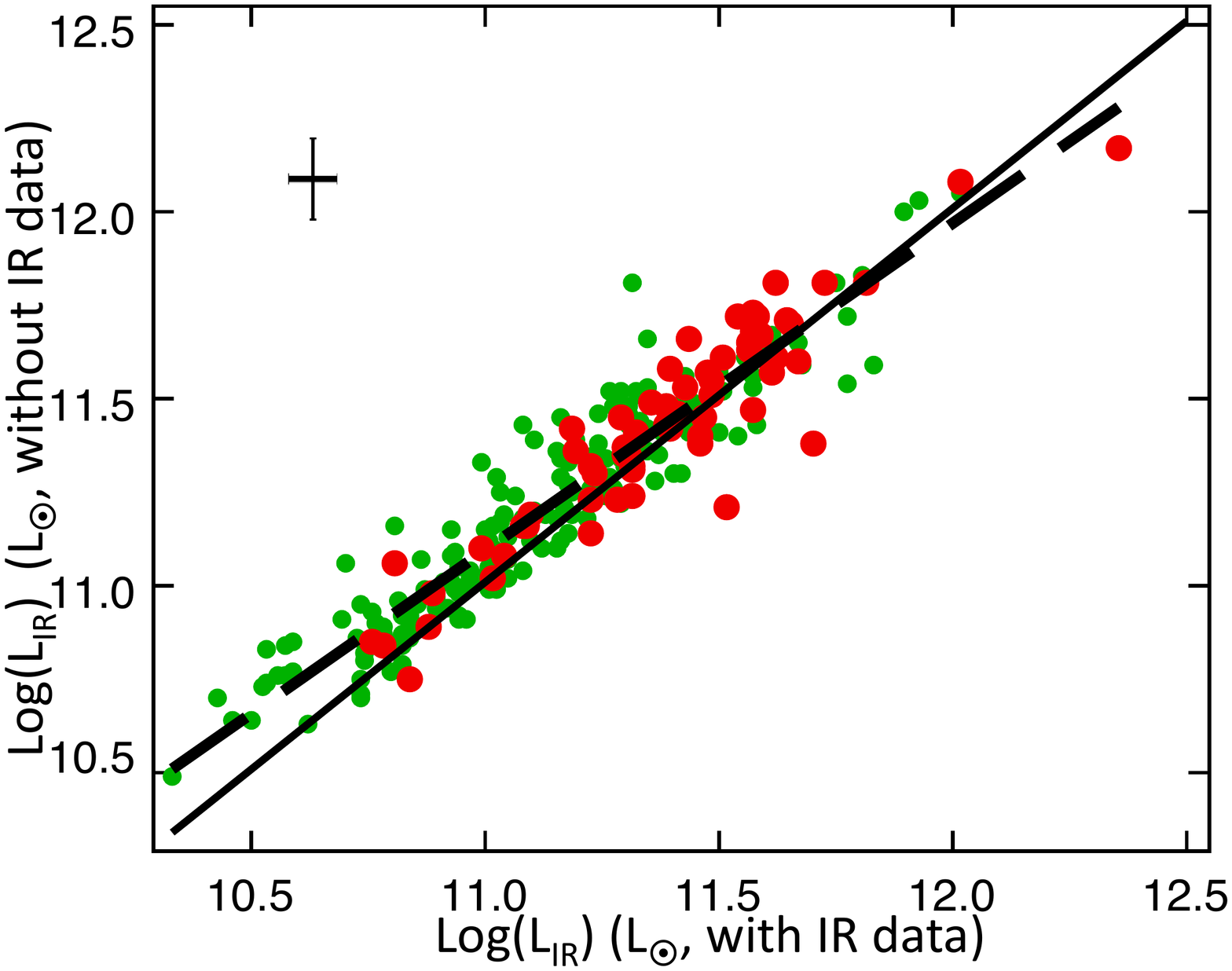}
\includegraphics[width=8.2cm]{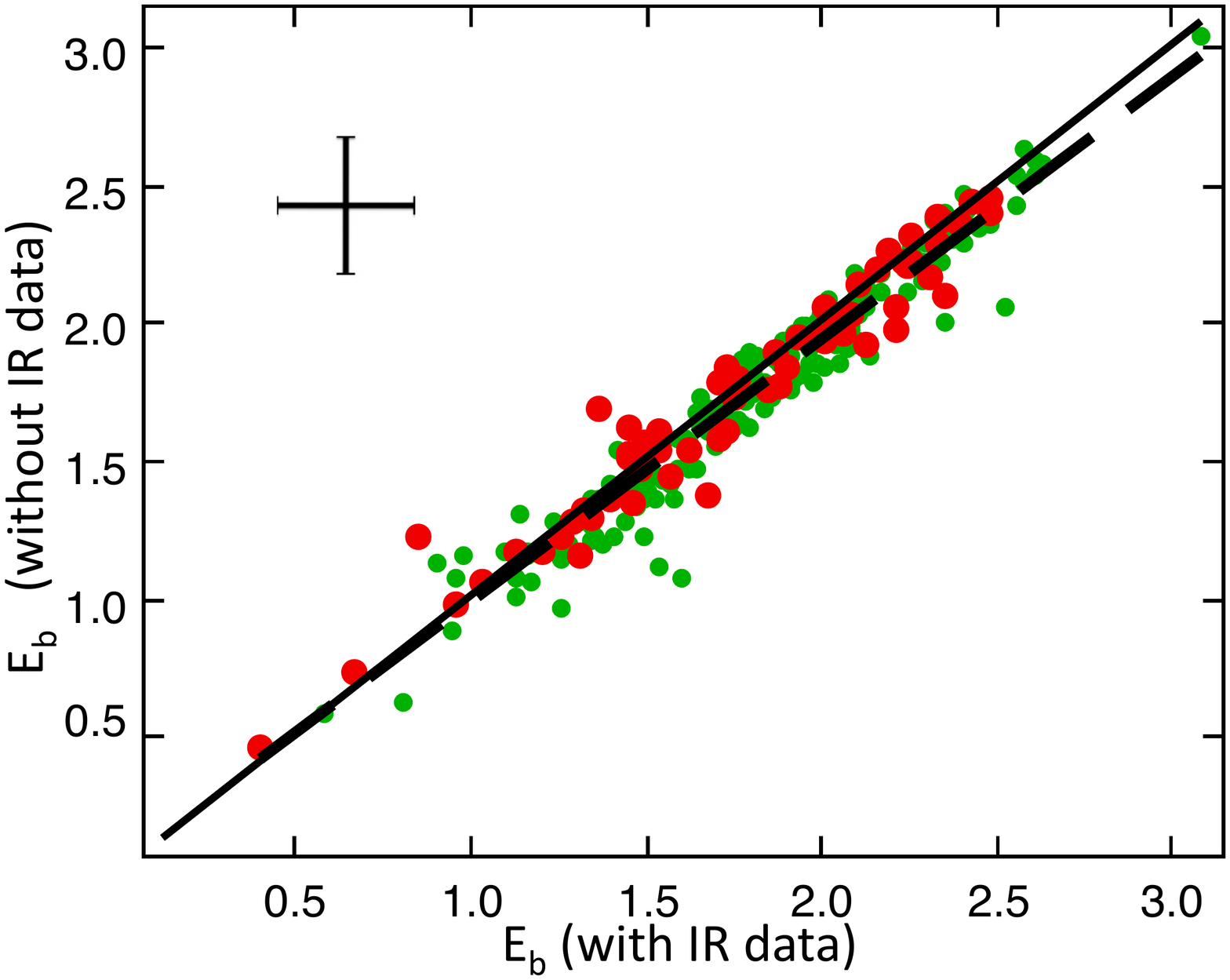}
\includegraphics[width=8.2cm]{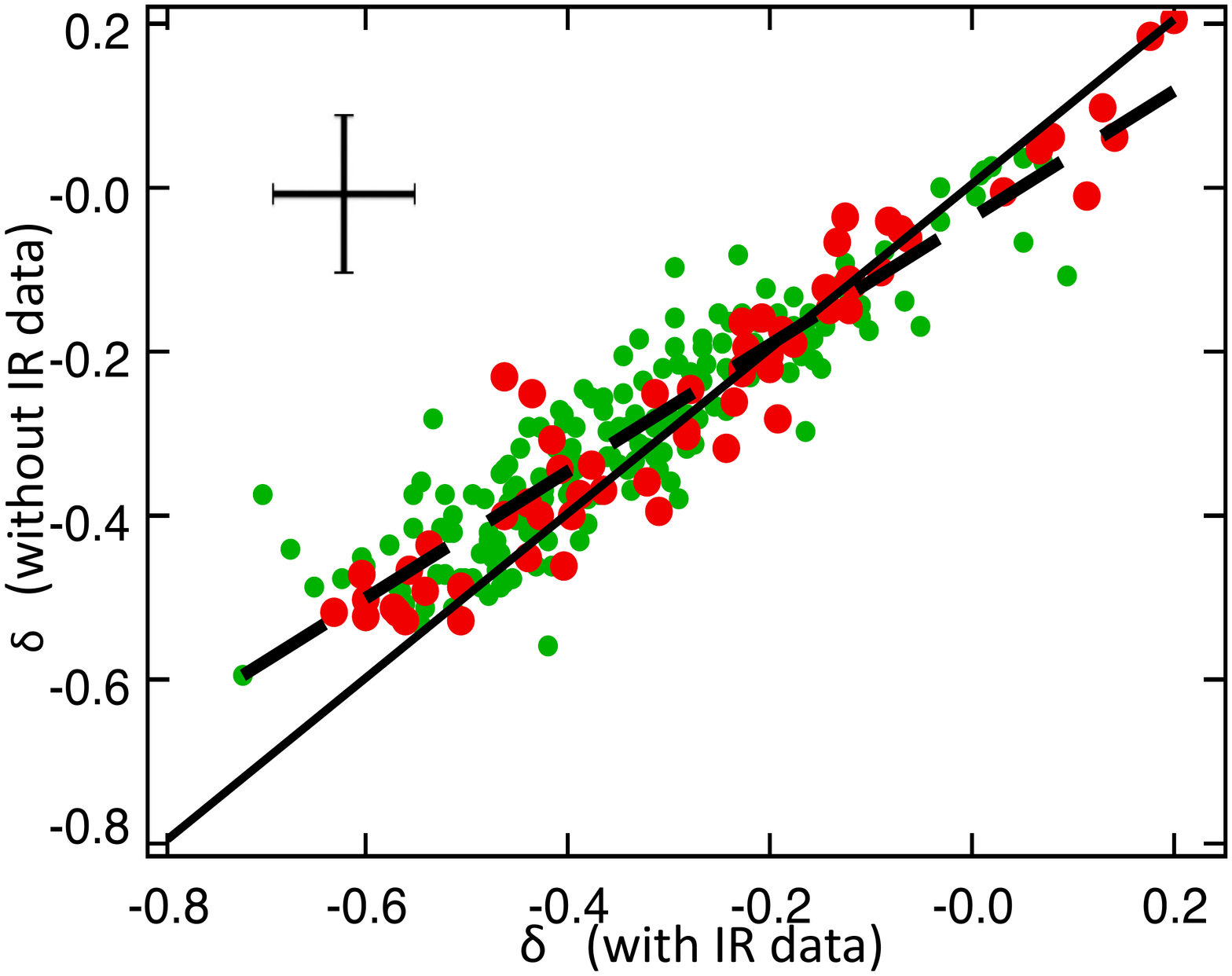}
\caption{Parameters estimated with available IR data (x-axis) are compared to those estimated without any IR data. Galaxies with only MIPS data are plotted with small green symbols, galaxies with MIPS and PACS detection with large red symbols. The green dashed line is the result of the linear regression and the black solid line is the 1:1 relation. Typical error bars (1\,$\sigma$) are plotted on the left corner of each plot.}
\label{fig:woIR}
\end{figure}

The results are summarised in Fig.~\ref{fig:mock} and Table~\ref{tab:correlation}. The sources detected with MIPS, with PACS and MIPS, and without any IR data are considered separately. For each parameter of interest in the present work, the correlation between the true and estimated values is shown in Fig.~\ref{fig:mock}. The correlation coefficient between both quantities is reported in Table~\ref{tab:correlation} along with the typical uncertainty. All parameters are better estimated when IR data are included. No clear difference is found when PACS data are used in addition to MIPS data to estimate $L_{\rm IR}$: this is due to the way we choose the IR fluxes of the artificial galaxies. Indeed, the determination of $L_{\rm IR}$ for artificial sources with MIPS data does not account for a variation of the IR templates, since artificial galaxies are built with only one IR SED ($\alpha = 2$), also used to fit them. Therefore, $L_{\rm IR}$ is well constrained with only one value. The SFR and $A_{\rm FUV}$ strongly depend on the determination of $L_{\rm IR}$ and are also well determined with only one IR (MIPS) data point. Without IR data, $L_{\rm IR}$ appears to be slightly overestimated, and as a consequence so are the SFR and $A_{\rm FUV}$. We will re-discuss the determination of $L_{\rm IR}$ in the next subsection. The input parameters related to the star formation history are not well constrained: previous studies have also confirmed the weak constraints placed on these parameters \citep{noll09b, giovannoli11, buat11a}. The parameters describing the dust attenuation curve, $E_{\rm b}$ and $\delta$, appear better constrained when IR data are available. The dynamical range of the estimated values is always lower than that of the true ones: the parameters are estimated as the mean of the PDF, which gives low weight to extreme values. In Sect.~\ref{sec:attcurve}, we will re-discuss this effect, which strongly affects the determination of $E_{\rm b}$.

\subsection{SED fitting with and without IR data}\label{sec:woIR}

It is important to check the robustness of the estimation of the most important parameters of this study against the availability of IR data. In complement to the analysis performed on artificial galaxies, we can also use the original sample with IR (MIPS or PACS data) to fit it with and without IR data and compare the estimated values of these parameters. We consider the sources detected by MIPS (290 galaxies) adding PACS data when available (for 76 galaxies). We run CIGALE on the SEDs of these galaxies without any IR data. In this case, only one IR SED is used with $\alpha = 2$ and $L_{\rm IR}$ is constrained by the obscuration of the UV-optical light. In Fig.~\ref{fig:woIR}, we compare the estimates of $A_{\rm FUV}$, $L_{\rm IR}$, $E_{\rm b}$, and $\delta$ with and without considering IR data. The agreement between the estimates of all parameters is very good with correlation coefficients equal to 0.95 for $A_{\rm FUV}$, $L_{\rm IR}$, and $E_{\rm b}$, and 0.92 for $\delta$. The SED fitting without IR data is found to slightly overestimate dust attenuation and IR luminosities for small values of these parameters, by 0.27\,mag for $A_{\rm FUV} = 1$\,mag and by 0.15\,dex for $\log(L_{\rm IR}) = 10.5~[L_{\odot}]$. It is qualitatively consistent with the analysis performed on artificial galaxies with an overestimate of 0.4\,mag for $A_{\rm FUV} = 1$\,mag and 0.20\,dex for $\log(L_{\rm IR}) = 10.5~[L_{\odot}]$ for artificial galaxies without IR data (Fig.~\ref{fig:mock}). The discrepancy between exact and estimated values of $A_{\rm FUV}$ and $L_{\rm IR}$ is larger for artificial galaxies without IR data, with a slope between the true and estimated values of $A_{\rm FUV}$ and $L_{\rm IR}$ significantly different from 1 (see Fig.~\ref{fig:mock}), than between these quantities estimated on real galaxies observed in IR and analysed with or without IR data (see Fig.~\ref{fig:woIR}). Most of the pseudo-galaxies without any IR data are intrinsically faint when compared to the galaxies detected in IR and plotted in Fig.~\ref{fig:woIR}. This difference together with the nature of the data (observed and artificial fluxes) can explain the relative discrepancies found in the two analyses performed here and in Sect.~\ref{sec:mock}. The determination of $E_{\rm b}$ is found to be robust against the inclusion or neglection of IR data, whereas $\delta$ is more uncertain without IR data and is overestimated for low values of the parameter (by 0.1~units for $\delta = -0.5$). The latter result was also obtained with the catalogue of artificial galaxies.

\subsection{Star formation history}\label{sec:SFH}

\begin{figure}
\centering
\includegraphics[width=9cm]{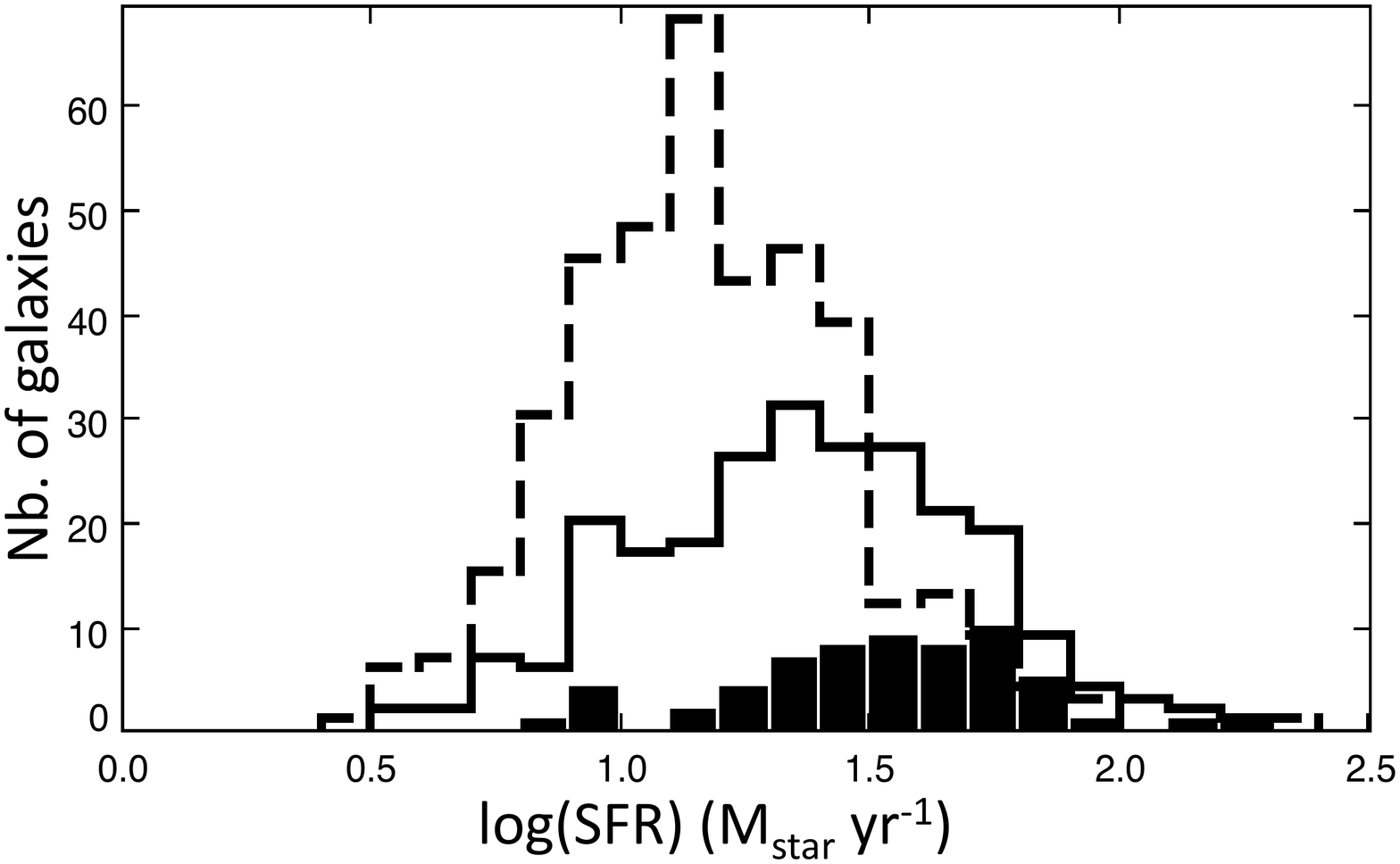}
\caption{Histogram of the distribution of the SFR obtained with SED fitting. The distribution with a dashed line corresponds to galaxies without IR detection, the solid line to galaxies detected by MIPS, and the filled histogram is obtained for galaxies detected by PACS.}
\label{fig:SFR}
\end{figure}

As shown in Sect.~\ref{sec:mock}, the detailed star formation history is not well constrained. The old stellar population is always found older than 1\,Gyr, from $\simeq 1.5$\,Gyr at $z \simeq 2$ to $\simeq 3$\,Gyr at $z \simeq 1$. The average mass fraction locked in the young stars ($f_{\rm ySP}$) is found to be of the order of $10\%$, which corresponds to galaxies with active star formation. In this work, we focus on the analysis of the UV part of the spectrum, which is dominated by the emission of the young stars. The average age of the young stellar component is 40\,Myr: a steady state in the production of the UV continuum is not reached \citep{leitherer99}. The distribution of the instantaneous SFR is shown in Fig.~\ref{fig:SFR}.

\section{The amount of dust attenuation}\label{sec:attenuation}

\begin{figure}
\centering
\includegraphics[width=9cm]{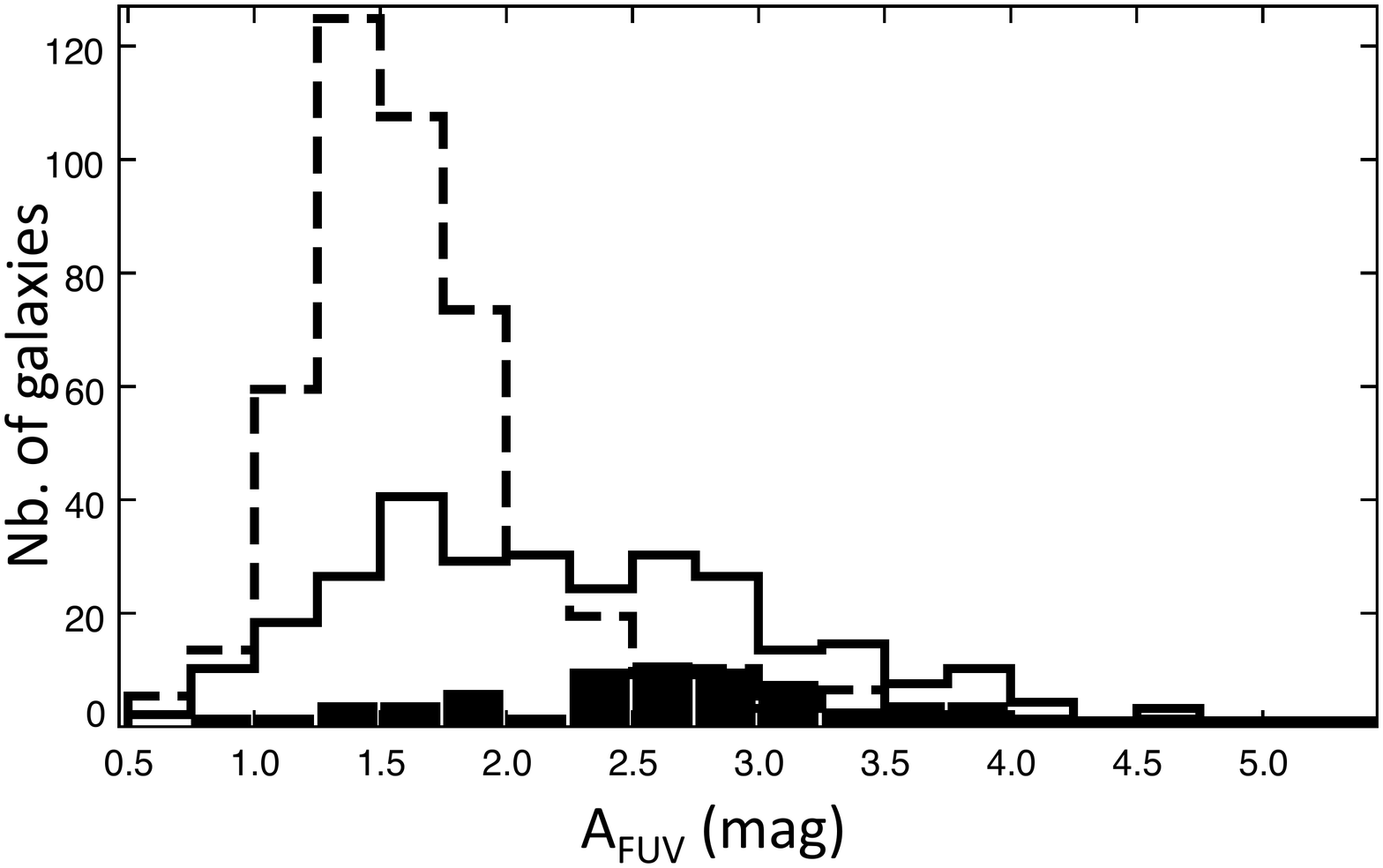}
\caption{Histogram of the dust attenuation in the FUV band. The distribution of $A_{\rm FUV}$ is plotted with a dashed line for galaxies without IR detection, a solid line traces galaxies detected by MIPS, and the filled histogram indicates galaxies detected by PACS.}
\label{fig:extinction}
\end{figure}

Several output parameters of the code quantify the amount of dust attenuation. In addition to the $V$ band attenuation of the young stellar population, the total amount of attenuation in the $V$ and FUV bands (for both stellar populations) are also calculated. The distribution of $A_{\rm FUV}$ is shown in Fig.~\ref{fig:extinction}. The values range from 0.5 to 5\,mag, the largest attenuations being found for galaxies detected with PACS. The mean dust attenuation is equal to 1.6, 2.2, and 2.6\,mag for the galaxies without IR data, with MIPS data only, and with MIPS and PACS data, respectively. Note that dust attenuation is likely to be slightly overestimated for galaxies without any IR detection (Sect.~\ref{sec:mock}).

The relative amount of attenuation between the young and the old stellar population, expressed in magnitudes, is measured by $f_{\rm att}$. As mentioned above, $f_{\rm att}$ is fixed to 0.5 for our baseline scenario, but we also performed runs with $f_{\rm att}$ as a free parameter allowed to take 3 values: 0.25, 0.5, and 1. This parameter is known to be ill-constrained \citep{noll09b,buat11a}. The value 1 is never chosen for the best-fit model and the mean value from the probability distribution function is found to be $0.52 \pm 0.06$. As a further check, we also ran CIGALE by also fixing $f_{\rm att}$ at 0.25 and 1. All results presented in this work are found unaffected by the exact value of this parameter. In particular, the distributions of $A_{\rm FUV}$, $L_{\rm IR}$, $M_{\rm star}$, $E_{\rm b}$, and $\delta$ are not significantly modified when $f_{\rm att}$ is varied.

\begin{figure}
\includegraphics[width=9cm]{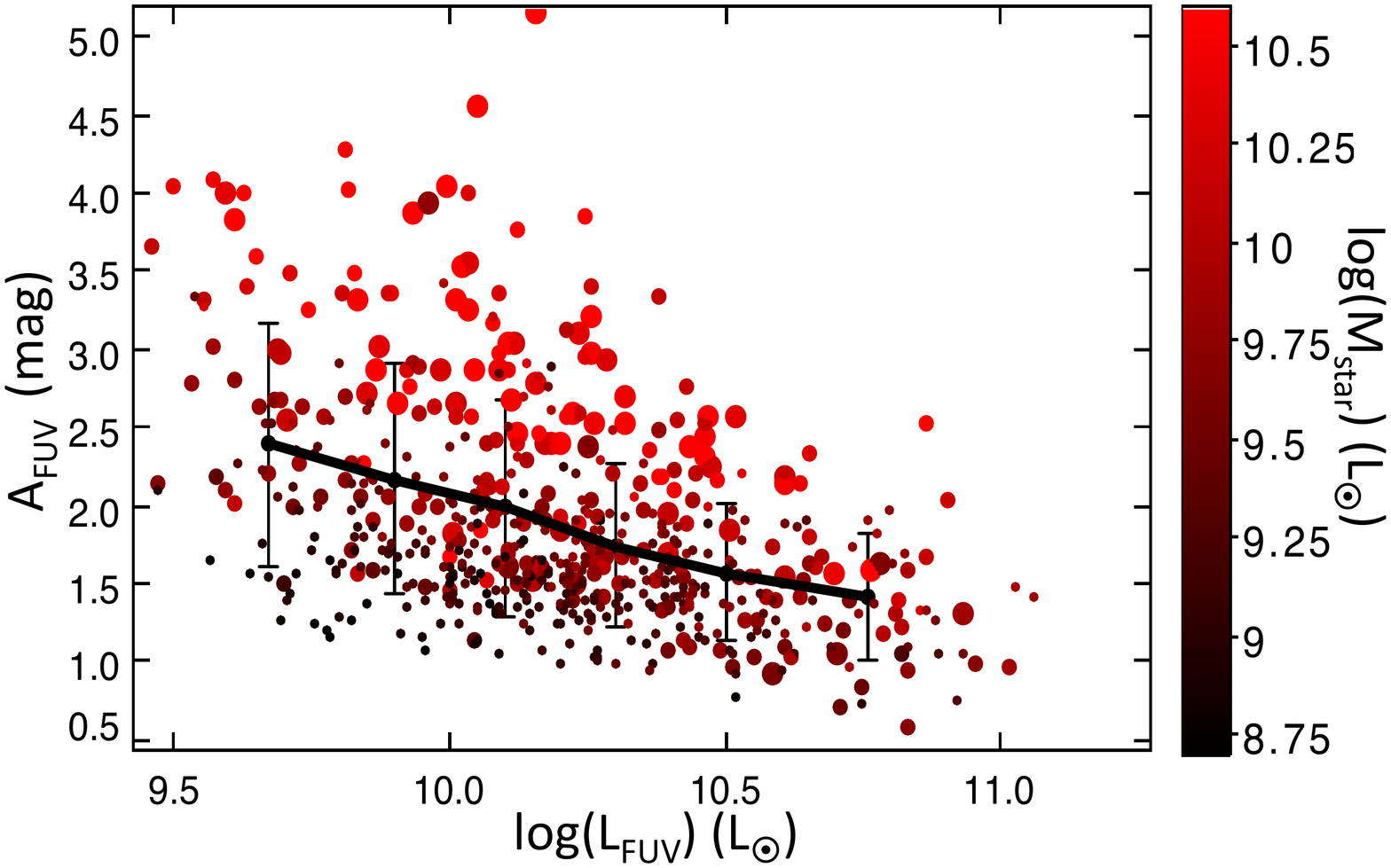}
\includegraphics[width=9cm]{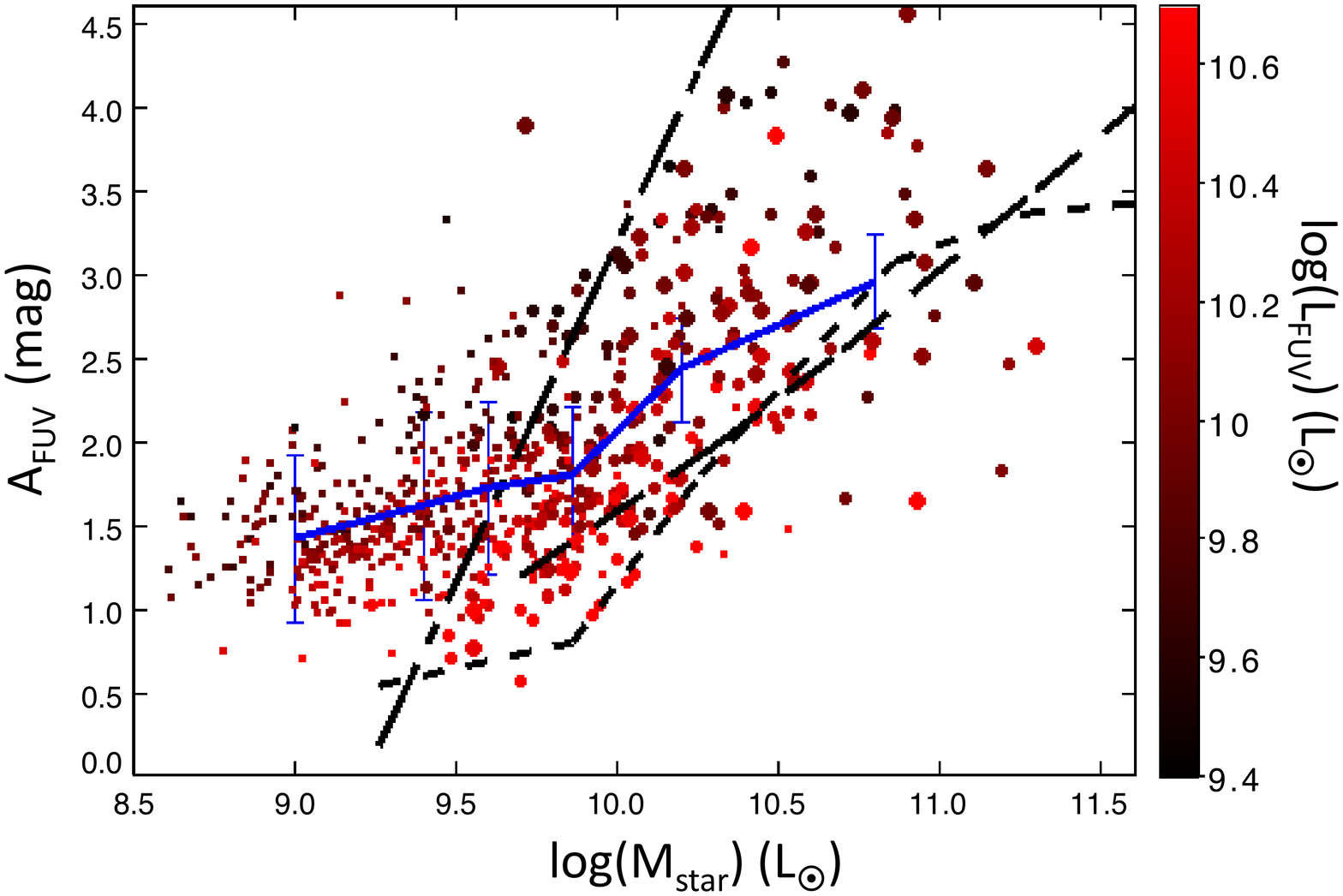}
\caption{Dust attenuation in FUV versus observed FUV luminosity and stellar mass. Galaxies with no IR data are plotted as small dots, those with MIPS data with medium dots, and those with MIPS and PACS detection with large dots. {\em Upper panel:} $A_{\rm FUV}$ versus $\log(L_{\rm FUV})$, colour coded with $\log(M_{\rm star})$. The average values of $A_{\rm FUV}$ per bin of $\log(L_{\rm FUV})$ are overplotted with their standard dispersion, calculated by accounting for the uncertainty on the determination of $A_{\rm FUV}$. {\em Lower panel:} $A_{\rm FUV}$ versus $\log(M_{\rm star})$, colour coded with $\log(L_{\rm FUV})$. The average values of $A_{\rm FUV}$ per bin of $\log(M_{\rm star})$ are overplotted with their standard dispersion (blue solid line). Relations from the literature and discussed in the text are overplotted as short dashed line (\citet{martin07}), dashed line (\citet{buat09}) at $z \simeq 1$, and dot-dashed line (\citet{panella09}) at $z \simeq 2$.}
\label{fig:Auv}
\end{figure}

In Fig.~\ref{fig:Auv}, $A_{\rm FUV}$ is plotted versus $L_{\rm FUV}$ and $M_{\rm star}$. There is a decrease of the upper envelope of the dust attenuation for large $L_{\rm FUV}$ in agreement with other findings based on UV selected samples at similar redshifts \citep{burgarella07,buat09, burgarella11,heinis12}. For a given UV luminosity, dust attenuation increases with stellar mass. Average values of $A_{\rm FUV}$ are calculated per bin of UV luminosity and are also reported in Fig.~\ref{fig:Auv}: in spite of a large dispersion, the mean value of $A_{\rm FUV}$ is found to decrease when $L_{\rm FUV}$ increases from $2.4 \pm 0.8$\,mag for $\log(L_{\rm FUV}) = 9.7~[L_{\odot}]$ to $1.5 \pm 0.4$\,mag for $\log(L_{\rm FUV}) = 10.7~[L_{\odot}]$, which corresponds to a slope -0.9 in Fig.~\ref{fig:Auv}. The trend still holds when only galaxies not detected in IR are considered with a less steep decrease (slope reduced to -0.6). The results are not modified if we account for the over-estimation of $A_{\rm FUV}$ for galaxies without an IR detection (cf. Sects.~\ref{sec:CIGALE} and \ref{sec:mock}). So the general variation seen in Fig.~\ref{fig:Auv} cannot be attributed only to the sources detected in IR, with MIPS or PACS. An inverse trend was reported at $z\sim 2$ for Lyman break galaxies by Reddy and collaborators \citep{reddy08,reddy10} (but see also \citet{burgarella11}). Interestingly, if we only consider galaxies with $z > 1.8$ to mimic a $z \simeq 2$ selection, our subsample contains galaxies more luminous than $10^{10}\,L_{\odot}$ and the distribution of $A_{\rm FUV}$ as a function of $L_{\rm FUV}$ is consistent with a flat one (although we do not exclude a slight decrease when $L_{\rm FUV}$ increases). \citet{reddy10} also have galaxies with $L_{\rm FUV} > 10^{10}\,L_{\odot}$, which can explain the apparent discrepancy between their analysis and the present work. The increase of the mean (and dispersion) of dust attenuation for galaxies fainter than $10^{10}\,L_{\odot}$ in UV is confirmed by \citet{heinis12} from a stacking analysis of Herschel data in the COSMOS field.

At higher redshift, from studies of the slope of the UV continuum, \citet{bouwens09} and \citet{lee11} also found a decrease of the dust attenuation with the absolute UV magnitude. However, \citet{finkelstein11} and \citet{dunlop12} found no significant variation of the UV slope with the absolute UV magnitude.

Dust attenuation is also found to be correlated with the stellar mass as already reported from the local universe up to high redshifts \citep{martin07,iglesias07,buat09,panella09,garn10,finkelstein11,sawicki12}. \citet{martin07} and \citet{buat09} studied UV selected samples up to $z \sim 1$. They reported variations of $A_{\rm FUV}$ as a function of the stellar mass or $K$-band rest-frame luminosity. The mean relations they found are shown in Fig.~\ref{fig:Auv} after some transformations. The relation of \citet{martin07} for $z = 1$ given in Fig.~\ref{fig:Auv} was obtained for a Salpeter IMF and the models of \citet{bruzual03}. We translate it to a Kroupa IMF by reducing the stellar masses by 0.24\,dex \citep{arnouts07}. The contribution of TP-AGB, included in the models of \citet{maraston05}, is assumed to further reduce the stellar mass by 0.2\,dex \citep{muzzin09}. \citet{buat09} proposed a relation between $\log(L_{\rm IR}/L_{\rm FUV})$, the rest-frame $K$-band, and the FUV luminosities, and did not find any trend with redshift. The relation reported in Fig.~\ref{fig:Auv} corresponds to a mean FUV luminosity $\langle\log(L_{\rm FUV})\rangle = 10.2~[L_\odot]$ representative of our current sample. The $K$-band luminosity is transformed in stellar mass using the calibration of \citet{arnouts07} modified by \citet{ilbert10} and $A_{\rm FUV}$ is deduced from $\log(L_{\rm IR}/L_{\rm FUV})$ following \citet{buat11a}. Although the slope of these correlations are consistent with what we find here, the average dust attenuation at a given stellar mass in these previous works is lower than our present estimate. We also compare our results to the relation obtained by \citet{panella09} at $z = 2$ by comparing SFRs from radio and UV measurements. We apply the corrections described above to translate their results to a Kroupa IMF, also adding the contribution of TP-AGB. Their relation is much steeper than the mean trend we find, leading to larger $A_{\rm FUV}$ at high $M_{\rm star}$, where IR data are available (i.e. when our estimates are the most reliable). Anyway, it is difficult from this comparison to draw a firm conclusion about the evolution of dust attenuation with redshift at a given stellar mass. Different methods were used to derive stellar masses and dust attenuation factors, and sample selections are also different. A UV selection like that used in this work is biased against high dust attenuation and stellar mass which might induce a shallower relation than for a mass- or SFR-selected sample. Moreover it is very difficult to estimate stellar masses, especially in blue, low-mass galaxies \citep{ilbert10}. Stellar masses and dust attenuation must be measured with the same method at all redshifts and within samples selected in the same way in order to be compared. Nevertheless, we can conclude that there is a net increase of dust attenuation at larger stellar masses and that the amount of dust attenuation also varies with the (uncorrected) UV luminosity.

\section{The attenuation curve}\label{sec:attcurve}

\subsection{Estimation of $E_{\rm b}$ and $\delta$}\label{sec:ebdelta}

\begin{figure}
\centering
\includegraphics[width=9cm]{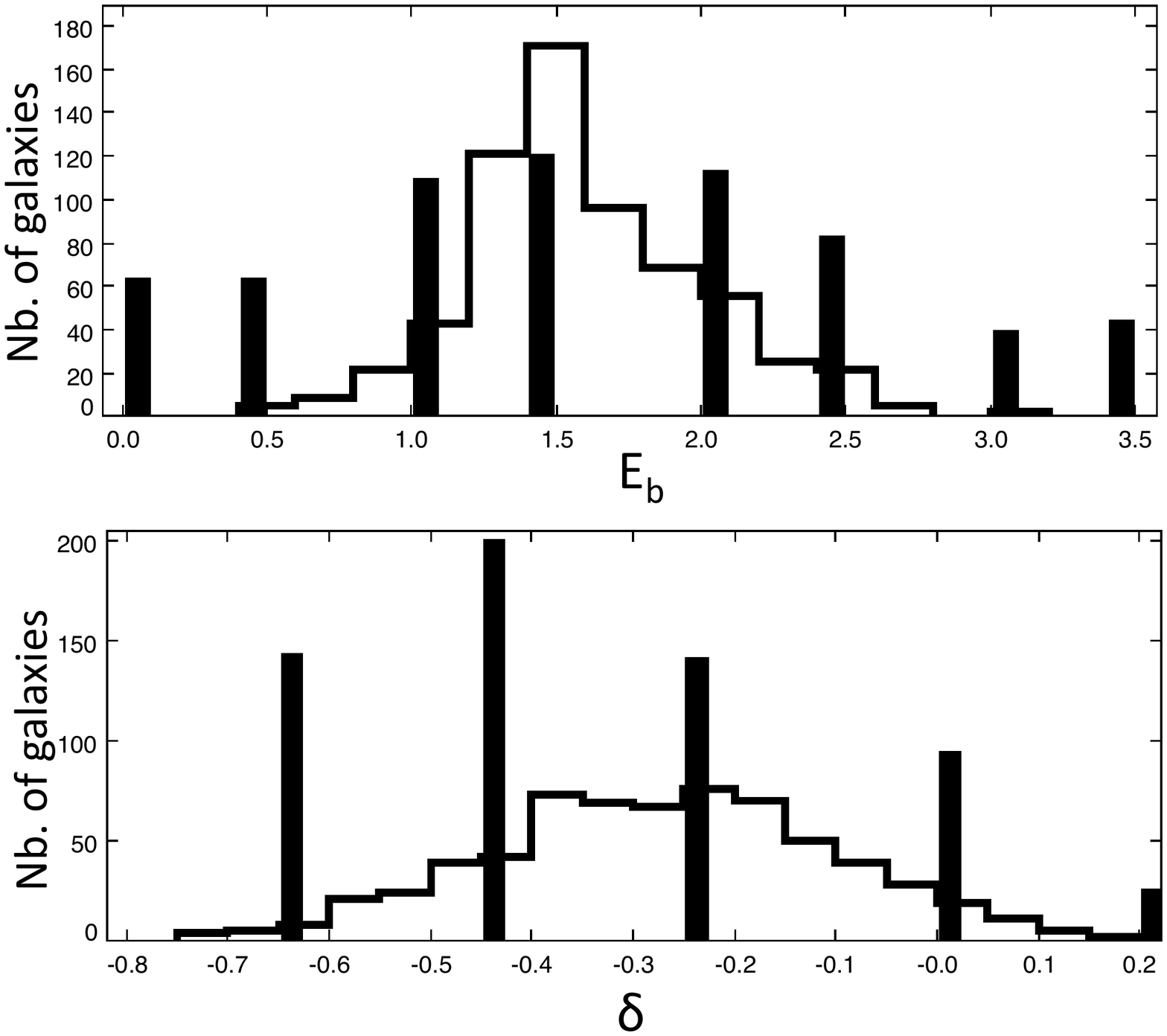}
\caption{Distribution of parameters of the attenuation curve. {\em Upper panel:} distribution of $E_{\rm b}$ values obtained for the best models (filled histogram) and from the analysis of the PDF (solid line). {\em Lower panel:} distribution of $\delta $ values obtained for the best models (filled histogram) and from the analysis of the PDF (solid line).}
\label{fig:histebdelta}
\end{figure}

\begin{figure}
\centering
\includegraphics[width=9cm]{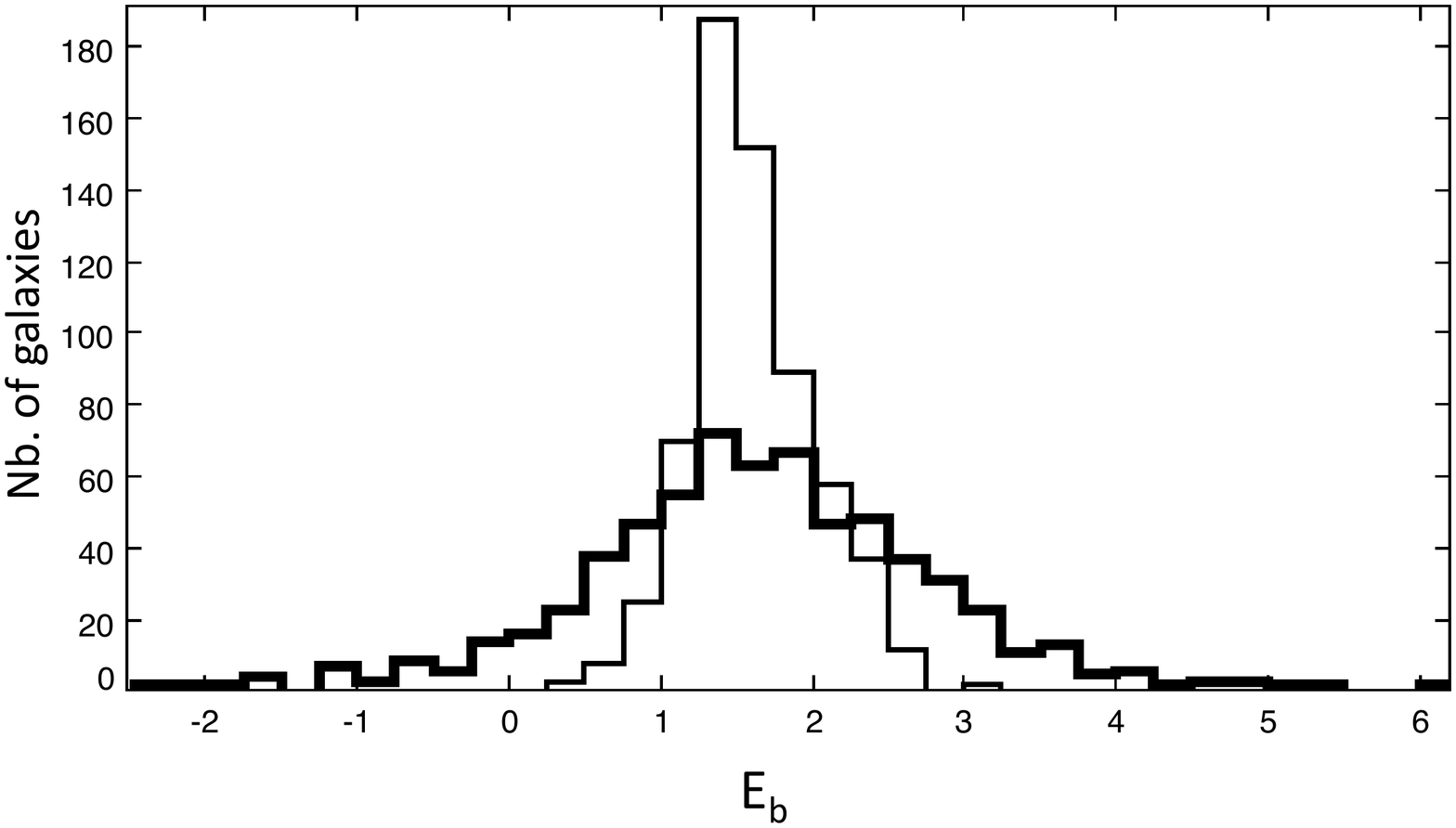}
\caption{Distribution of $E_{\rm b}$. Thin line: $E_{\rm b}$ estimated as the mean of the PDF. Thick line: $E_{\rm b}$ values corrected for edge effects (see text for details). The regressions used for the corrections are: $E_{\rm b} ({\rm SED fitting}) = 0.26 E_{\rm b} ({\rm model}) + 1.08$, $E_{\rm b} ({\rm SED fitting}) = 0.47 E_{\rm b} ({\rm model}) + 0.91$, and $E_{\rm b} ({\rm SED fitting}) = 0.56 E_{\rm b} ({\rm model}) + 0.74$ for the samples without IR detection, with MIPS detections only, and with MIPS and PACS detections,respectvely.}
\label{fig:histcorr}
\end{figure}

The dust attenuation curve is described by the parameters $E_{\rm b}$ and $\delta$ (Eqs.~\ref{eq:attlaw} and \ref{eq:bump}).

\begin{figure}
\centering
\includegraphics[width=9cm]{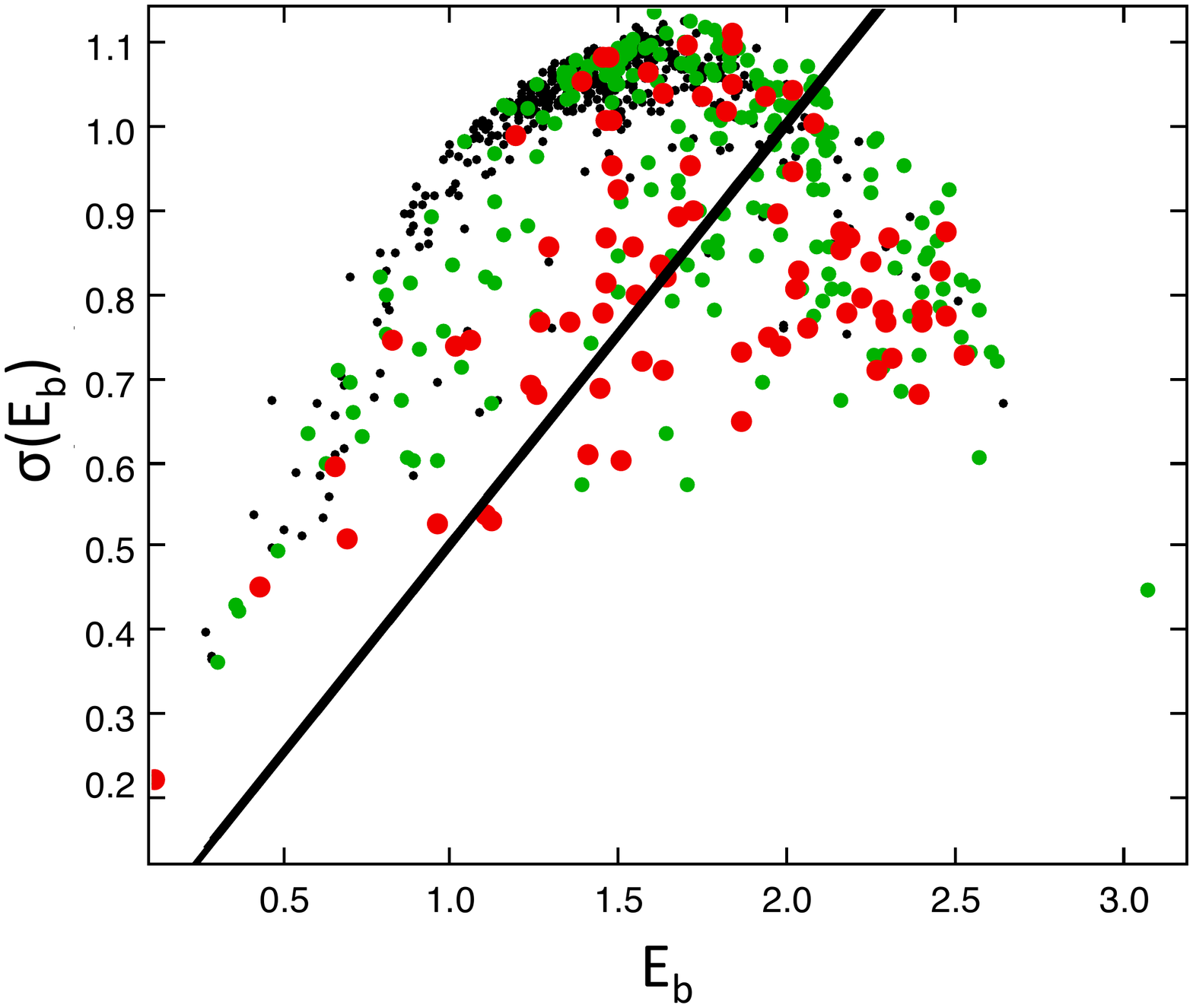}
\includegraphics[width=9cm]{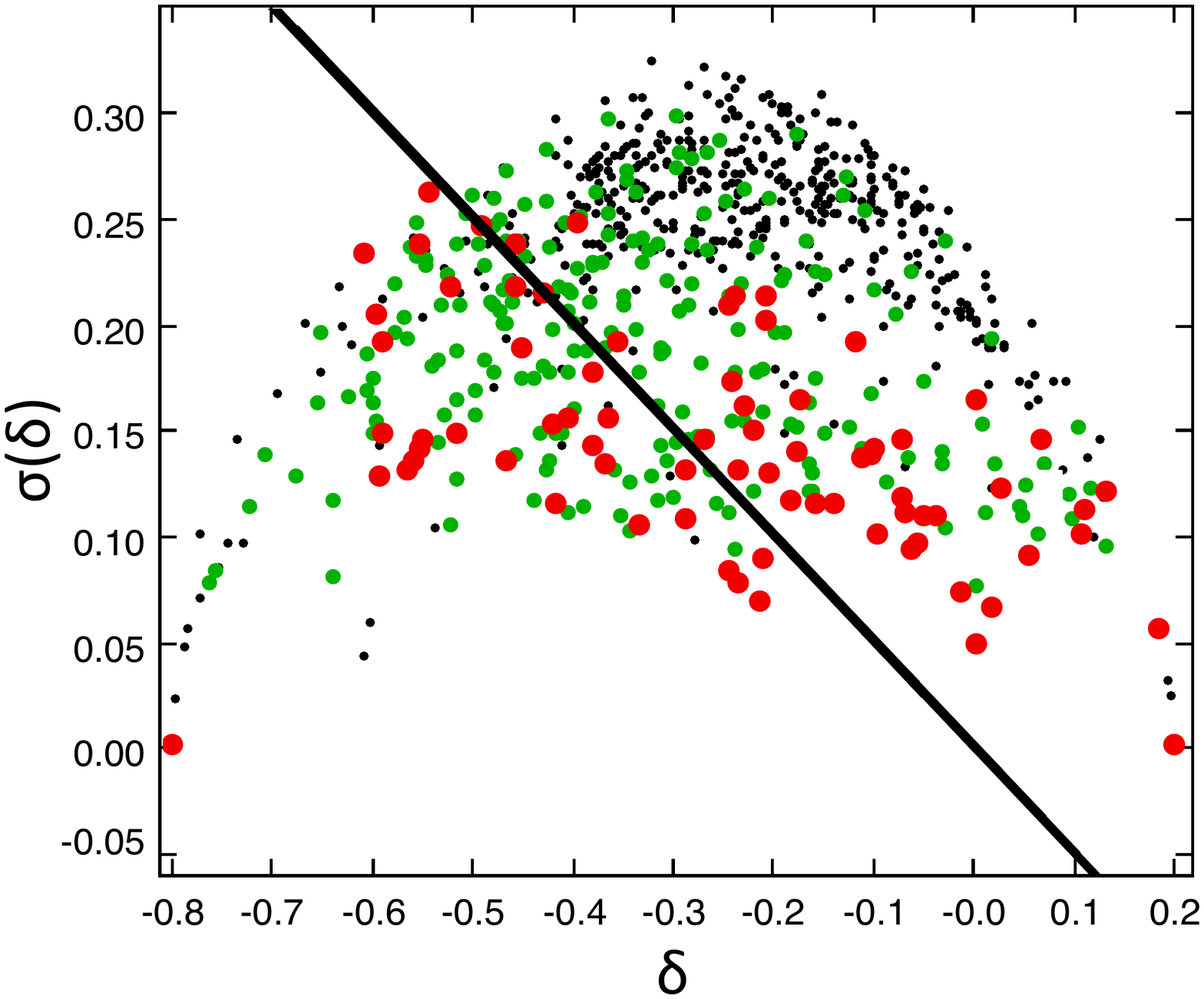}
\caption{Mean (x-axis) and standard deviation (y-axis) for the estimated values of $E_{\rm b}$ (upper panel) and $\delta$ (lower panel). In the upper panel, galaxies with a secure ($> 2\,\sigma$) detection of the bump are located to the right side of the solid line. In the lower panel, galaxies with an attenuation curve confirmed to be steeper than the Calzetti one are located to the left side of the solid line (see text for details). The symbols are the same as in Fig.~\ref{fig:mock}.}
\label{fig:ebdelta}
\end{figure}

The distributions of these parameters for the best models, estimated as the mean of the PDF, are plotted in Fig.~\ref{fig:histebdelta}. The extreme values of $\delta$ are not commonly chosen for the best model obtained by $\chi^2$ minimisation. However, it is not the case for $E_{\rm b}$ for which the null value is well represented in the best fits. By definition, only positive values of $E_{\rm b}$ are possible and the output value, given by the code as the mean of the PDF, will be overestimated when the exact value is close to 0. The truncation of the PDF, when its mean value is close to the extreme input values of the parameter distribution, induces an over- or underestimation of the parameter (see Fig.~\ref{fig:mock} and Sect.~\ref{sec:mock}). The amplitude of the effect depends on the exact shape of the PDF. \citet{noll09b} discussed the case of a Gaussian distribution with a given standard deviation: for a standard dispersion of 1 unit, typical of our results for $E_{\rm b}$ (see below and Fig.~\ref{fig:ebdelta}), all the values of $E_{\rm b}$ lower than 1.2 are affected and the correction ranges from 0.6 for $E_{\rm b} = 0$ to 0 for $E_{\rm b} = 1.2$. Another way to quantify the effect, which is less dependent on the shape of the PDF, is to use the correlations obtained for the artificial galaxies between the exact and the measured values of $E_{\rm b}$. In Fig.~\ref{fig:mock}, we performed linear regressions, splitting the mock sample in objects without IR detections, those detected by MIPS, and objects detected by PACS. The exact value of $E_{\rm b}$ ($E_{\rm b}$(model)) was chosen as independent variable. The resulting regression formulae are applied to the observed sample to correct the values of $E_{\rm b}$ obtained from the PDF of the SED modelling. The results are plotted in Fig.~\ref{fig:histcorr} and the regression formulae are given in the figure caption. As expected, the distribution of the corrected values is found to be broader. Galaxies with unphysically negative values for $E_{\rm b}$ represent 6$\%$ of the sample, whereas all negative values are consistent with a null value of $E_{\rm b}$ at the one sigma level. The mean value for the entire sample remains unchanged (see below).

To further check the parameter ranges and the reliability of the results, we performed an additional run with a very large range of values for $E_{\rm b}$ and $\delta$: $-4 < E_{\rm b} < 4$ and $-0.9 < \delta < 0.9$ with a loose sampling to limit the computation time. Negative values of $E_{\rm b}$ are found for only 3$\%$ of the best models. The fraction reaches 16$\%$ for the mean value of the PDF, but except for one object, all these negative mean values are consistent with $E_{\rm b} = 0$ at a one sigma level. The distribution of $\delta$ is very similar to that obtained with the parameters of Table~\ref{tab:parameters}. This check confirms the results found with the baseline run and the reliability of the range of values found for $E_{\rm b}$ and $\delta$.

From an analysis of the mean and standard deviation of each PDF, we can measure the fraction of galaxies for which the bump is securely measured. Mean values and standard deviations for $E_{\rm b}$ and $\delta$ are plotted in Fig.~\ref{fig:ebdelta}. We adopt the same criterion as in paper~I to select sources with a secure detection of the bump as well as those with $\delta$ confirmed to be negative, which corresponds to a probability of at least 95$\%$ for $E_{\rm b} > 0$ and $\delta < 0$. Assuming a Gaussian distribution with mean and standard deviation given by the fit, it corresponds to $E_{\rm b} - 2 \sigma_{E_{\rm b}} > 0$ and $\delta + 2 \sigma_{\delta} < 0$. Using this criterion, 20$\%$ of the whole sample is exhibiting a secure bump. The percentage reaches 40$\%$ (resp. 47$\%$) for  MIPS (resp. PACS) detections. If we apply the corrections for the truncation of the PDF discussed above, the fraction of secure detections of the bump increases to 35$\%$ for the whole sample and to 54$\%$ (resp. 50$\%$) for MIPS (resp. PACS) detections. A similar percentage (20$\%$) of the whole sample shows a secure negative $\delta$. The percentage is 41$\%$ for galaxies detected with MIPS and 43$\%$ for those detected by MIPS and PACS. Note that only four SEDs are fitted with a $\delta$ value found securely positive (with $\delta - 2 \sigma_{\delta} > 0$. It is worth noting that the error on the estimation of $E_{\rm b}$ and $\delta$ is lower when the galaxies are detected with MIPS (or PACS).

From the above discussion, it is clear that we must be cautious when reporting mean values for $E_{\rm b}$ or $\delta$. The mean amplitude of the bump for the whole sample is $\langle E_{\rm b} \rangle = 1.6 \pm 0.4$. Applying the corrections for the truncation at the edge of the parameter range (see above) only increases the dispersion but not the mean value ($\langle E_{\rm b} \rangle = 1.6 \pm 1.1$). The mean amplitude is increased to $\langle E_{\rm b} \rangle = 1.8$ for galaxies detected in IR. These values are larger than those obtained in paper~I. The difference comes from the range of initial values adopted for $E_{\rm b}$ and the estimation of the parameter as the mean of the PDF. In paper~I, this range was $0 < E_{\rm b} < 2$, based on the fact that values larger than 2 were never taken for the best models. In the present work, we extend the range of values up to 3.5 (corresponding to the MW value) to minimise the impact of the truncation of the PDF, and because larger values than 2 are also obtained for some of the best models (Fig.~\ref{fig:ebdelta}). While the distribution of values based on $\chi^2$ minimisation are similar (same mean and median: $E_{\rm b} = 1.5$), the values estimated from the PDF are larger in the present case and were likely to be underestimated in paper~I. We ran the code with the parameters defined in the present work for the sample of galaxies used in paper~I and obtained $\langle E_{\rm b} \rangle = 1.6 \pm 0.3$.

The mean value $\langle E_{\rm b} \rangle = 1.6$ corresponds to 45$\%$ of the bump amplitude of the MW extinction curve and is similar to that of the LMC supershell region \citep{fitz07, gordon03}. The large dispersion we find is likely due to the combination of uncertainties in the determination of the bump as well as due to intrinsic differences among galaxies and different environments within a galaxy.

\begin{figure}
\centering
\includegraphics[width=9cm]{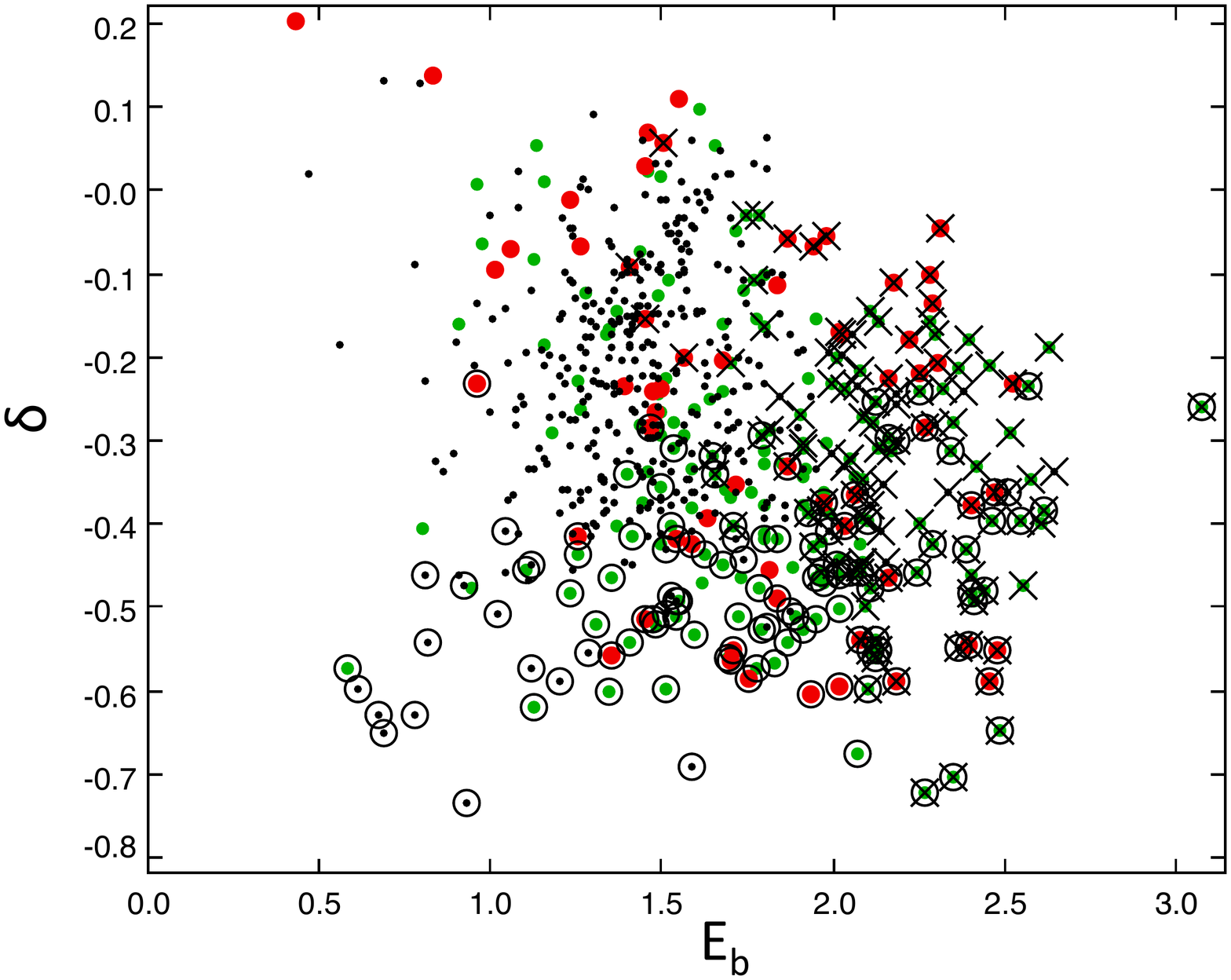}
\caption{Slope of the attenuation curve versus amplitude of the bump: symbols are the same as in Fig.~\ref{fig:mock}. Galaxies with a bump detected with a probability higher than 95$\%$ are overplotted with crosses and those with secure negative values of $\delta$ at the same level of confidence are inscribed in a large circle (see text for details).}
\label{fig:Eb-delta}
\end{figure}

The mean value of $\delta$ for the whole sample is $\langle \delta \rangle = -0.27 \pm 0.17$. No significant difference is found when galaxies with and without IR data are considered separately. The value found in paper~I was less negative: running CIGALE on the first sample with the present parameters gives $\langle \delta \rangle = -0.22 \pm 0.18$ compared to $\langle \delta \rangle = -0.13 \pm 0.12$ in paper~I. The difference between both estimates indicates the uncertainty: modifying the range of $E_{\rm b}$ is likely to imply a difference in the determination of $\delta$. The mean value found here ($ \langle \delta \rangle = -0.27$) corresponds to an increase in UV similar to that of the LMC extinction curve (Sect.~\ref{sec:Auv-beta}). We do not find any trend between $E_{\rm b}$ and $\delta$ (Fig.~\ref{fig:Eb-delta}).

\subsection{Differential slope and $E_{\rm b}$}\label{sec:Eb-deltabeta}

\begin{figure}
\centering
\includegraphics[width=9cm]{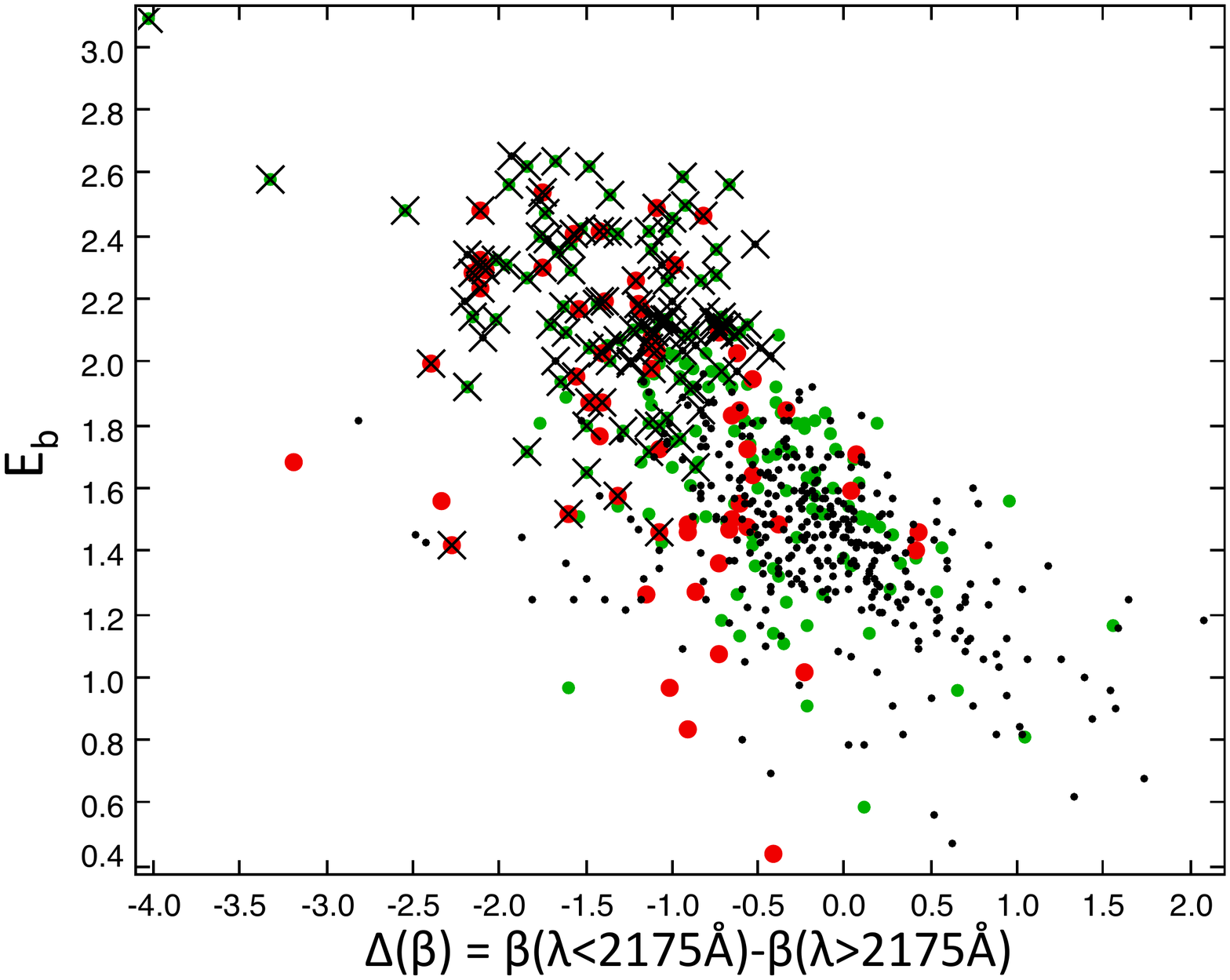}
\caption{Amplitude of the bump of the attenuation curve versus the observed variation of the slope of the UV continuum across the bump. The symbols are the same as in Fig.~\ref{fig:Eb-delta}.}
\label{fig:Eb-deltabeta}
\end{figure}

The imprint of the UV bump in attenuated spectra is expected to lead to a change of the UV slope across the UV bump. Following \citet{noll05}, who defined a spectral index $\gamma_{34}$, we split the photometric data in two subsamples corresponding to $\lambda < 2175\,\AA$ and $\lambda > 2175\,\AA$. We restrict the redshift interval from 1.1 to 1.8 in order to have at least five data points in each subsample. We calculate the values of $\beta$ corresponding to each wavelength range and subtract the two values with a linear fit: $\Delta\beta = \beta (\lambda < 2175\,\AA) - \beta (\lambda > 2175\,\AA)$. The standard error on this differential slope, obtained by adding quadratically the errors on each slope, is of the order of 0.7, i.e. quite large.

A clear trend is found in Fig.~\ref{fig:Eb-deltabeta}: all the galaxies with a secure bump detection exhibit a negative value of $\Delta(\beta)$. In order to have reference values, $\Delta(\beta)$ is also calculated on the SEDs of the best models using the same photometric bands as for the real data. Models without bump in the attenuation curve correspond to $\Delta(\beta) \simeq 0.5$. This is consistent with Fig.~\ref{fig:Eb-deltabeta} if we account for the large error found for $\Delta(\beta)$ and the overestimation of $E_{\rm b}$ for low values of this parameter (Sect.~\ref{sec:ebdelta}). Given all these uncertainties, this comparison has to be taken with care. However, our criterion $E_{\rm b} - 2 \sigma_{E_{\rm b}} > 0$ to select a secure detection of the bump appears to be valid and even conservative, since it corresponds systematically to negative values of $\Delta(\beta)$.

\subsection{Discussion}\label{sec:discussion}

We can summarise the results obtained on the attenuation curve for our galaxy sample: a large range of values is found for the amplitude of the bump and the slope of the curve. Since this dispersion is not significantly reduced by neglecting galaxies with the most uncertain values (see Fig.~\ref{fig:ebdelta}), we can state that these variations probably reflect a real diversity of attenuation curves. These results imply that SED fitting codes must allow for variations of the attenuation curve to fit the UV continuum of galaxies at all redshifts and retrieve reliable attenuations \citep{ilbert09}.

\begin{figure*}
\centering
\includegraphics[width=20cm]{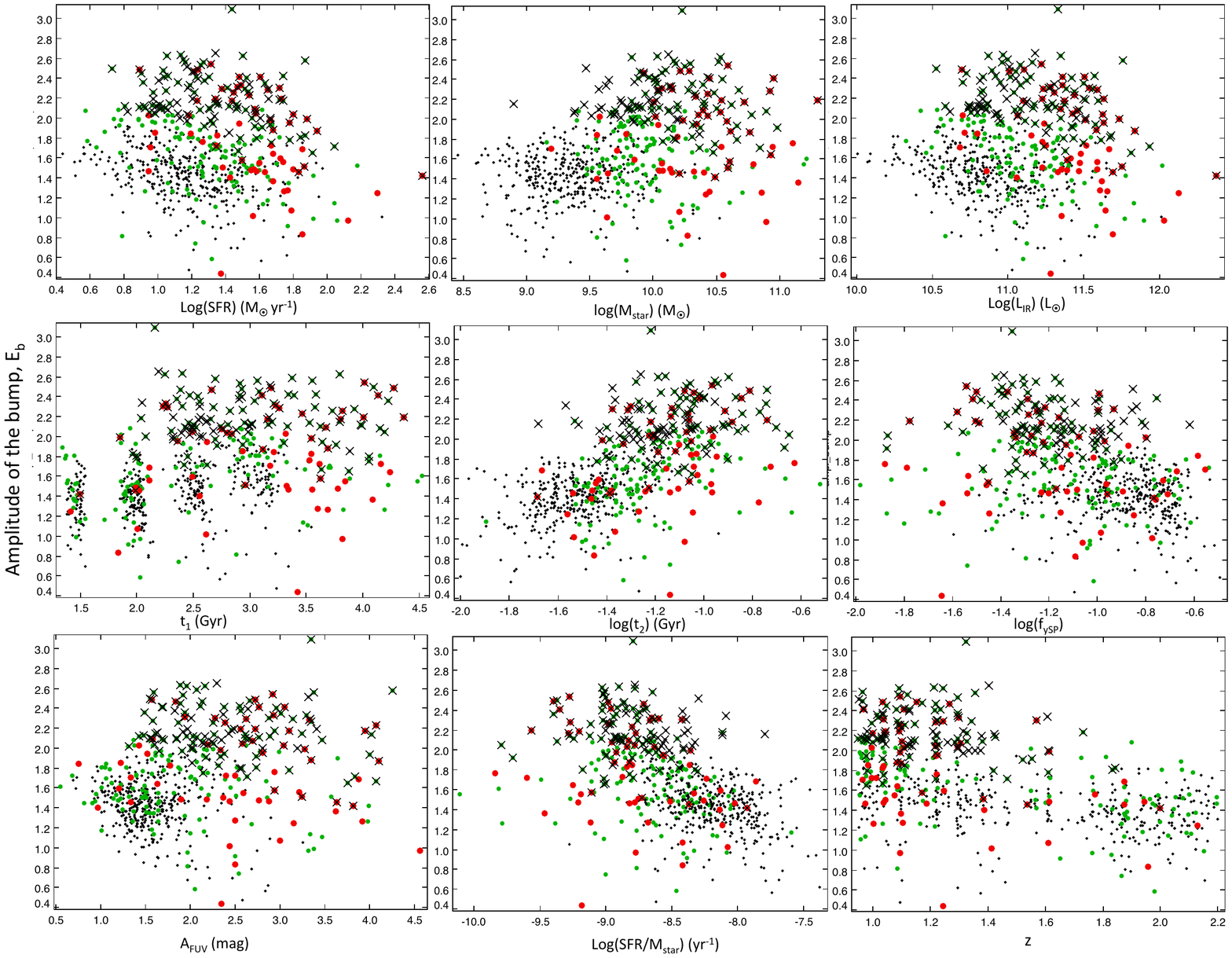}
\caption{Variation of the bump amplitude $E_{\rm b}$. Symbols are the same as in Fig.~\ref{fig:mock} with crosses overplotted on sources with a secure detection of the bump. From left to right and top to bottom: $E_{\rm b}$ is plotted versus $\log({\rm SFR})$, $\log(M_{\rm star})$, $\log(L_{\rm IR})$, the age of the old stellar population $t_1$ (Gyr), the age of the young stellar population $\log(t_2)$ (Gyr), the mass fraction of the young stellar population $\log(f_{\rm ySP})$, $A_{\rm FUV}$ (mag), the specific SFR $\log({\rm SFR}/M_{\rm star})$, and the redshift $z$.}
\label{fig:Eb-var}
\end{figure*}

The presence of a bump is confirmed in at least 20$\%$ of the sources and up to 35$\%$ if we account statistically for the truncation of the PDF of this parameter. A steeper attenuation curve than that of \citet{calzetti00} is also robustly found in 20$\%$ of our galaxies. \citet{noll09a} confirm the presence of a bump in $\sim 30\%$ of their sources, which is consistent with our analysis. The amplitude of the bump is plotted versus the parameters deduced from the SED fitting in Fig.~\ref{fig:Eb-var}. No correlation is found with SFR or $L_{\rm IR}$ and very weak ones may exist with $A_{\rm FUV}$ (correlation coefficient $R = 0.25$) or $M_{\rm star}$ ($R = 0.36$). $E_{\rm b}$ is found to be lower for younger stellar populations, higher fraction of young stars, high specific SFRs (defined as SFR/$M_{\rm star}$) and higher redshifts. To check the robustness of these trends against SED fitting uncertainties for galaxies not detected in IR, we applied corrections to $\log({\rm SFR})$, $\log(M_{\rm star})$, $\log(L_{\rm IR})$, $A_{\rm FUV}$, and $E_{\rm b}$ based on the regression lines plotted in Fig.~\ref{fig:mock} and found no significant modification (not shown here). \citet{noll09a} also found that galaxies with evidence of a 2175\,$\AA$ feature host older stellar populations than galaxies lacking an evident bump.

The decrease of the bump amplitude in galaxies forming stars actively could be related to a destruction of bump carriers in intense radiation fields or in supernovae shock fronts. Galaxies with high SFRs are recognised to be more compact than galaxies populating the main sequence defined by the SFR-$M_{\rm star}$ relation. The PAH molecules detected in the IR SEDs of the former galaxies exhibit lower equivalent widths \citep{elbaz11} and these molecules are often proposed to be at the origin of the UV bump \citep{xiang11}. A deficiency in the production of bump carriers can also be invoked to explain the decrease of $E_{\rm b}$ in small galaxies, forming stars actively. Most bump carriers or the pre-products of them should be produced by carbon-rich AGB stars, which appear about 300\,Myr after the onset of the star formation. Therefore, compact, low mass galaxies with a short but intense star formation history are not good places for the formation of a strong UV bump.

Conversely, radiative transfer models for a clumpy medium, mixing of dust and stars, and dust-clearing effects for older stellar populations predict a decrease in the amplitude of the bump and the effective attenuation even if the extinction law exhibits a large bump \citep[e.g.][]{witt00,pierini04,panuzzo07,noll07}. If galaxies with such attenuation properties significantly contribute to the sample, the fraction of galaxies with dust properties that allow for the formation of a strong UV bump should be distinctly higher than observed. Models assuming a clumpy distribution of stars and dust also predict a general flattening of the attenuation curve, although an age-selective attenuation acts in steepening the attenuation curve \citep{inoue06,panuzzo07}. A patchy distribution of the dust is needed to obtain an attenuation curve as grey as that of \citet{calzetti00} starting with a SMC-like extinction curve \citep{witt00, charlot00}. The mean attenuation curve found for our sample galaxies is steeper than the \citet{calzetti00} curve: if a significant flattening of the attenuation curve is at work, then the extinction curve must be very steep in our galaxies. The dust grain size distribution plays a major role in the general shape of the extinction curve and a steep increase in UV, which is attributed to a deficit of large grains \citep[e.g.][]{cartledge05}, is not unlikely in high-redshift galaxies forming stars very actively.

On the other hand, our sample is characterised by UV bump strengths and attenuation curve slopes close to those typically observed for the extinction curve in the LMC supershell region. \citet{noll09a} show that the typical bump widths of luminous star-forming galaxies at these redshifts also agree with the mean value of the LMC supershell region. Hence, the LMC supershell region may be a model region for high-redshift galaxies with significant UV bumps. In that case, the attenuation curves have to be similar to the extinction curves and this can also be achieved if the dust tends to show a screen-like distribution. This assumption is supported by the observation of strong interstellar absorption lines in galaxies with significant UV bump \citep{noll05,noll09a}, which can be interpreted by a high covering fraction of young massive stars by dusty gas clouds. Screen-like dust topologies can be produced by large-scale outflows, which is a reasonable assumption for vigorously star-forming high-redshift galaxies.

\section{Dust attenuation and UV slope}\label{sec:att_slope}

The slope $\beta$ of the UV continuum has been widely used to measure dust attenuation in galaxies since the works of Calzetti, Meurer, and collaborators. They showed that the intrinsic value of this slope does not vary a lot as long as a high star formation activity is maintained in galaxies \citep{leitherer99,calzetti01}: any deviation from this value is assumed to be caused by dust attenuation. This was confirmed by the correlation between the observed slope $\beta$ and the amount of dust attenuation measured by $L_{\rm IR}/L_{\rm FUV}$ found for local starbursts \citep[e.g.][]{Meurer99, overzier11}. Although the situation is more complex for non-starbursting galaxies \citep[e.g.][and references therein]{burgarella05, boissier07, hao11,boquien12}, this relation seems to hold for high-redshift systems forming stars intensively, such as Lyman break or BzK galaxies \citep{daddi07, reddy08,magdis10}.

We have a reliable measure of $\beta$ thanks to the intermediate bands included in the analysis. The presence of a bump could modify substantially the value of $\beta$ measured with broad-band data if they cover the bump area \citep{burgarella05,buat11b}. However, our measure of $\beta$ is insensitive to the presence of the bump (cf. Sect.~\ref{sec:UV}). We restrain the following study to galaxies detected in IR. Even if the determination of parameters related to dust attenuation is satisfactory without IR data (Sect.~\ref{sec:SED-fitting}), the $L_{\rm IR}/L_{\rm FUV}$-$\beta$ diagnostic is very sensitive to variations of the dust attenuation and star formation histories. We prefer to restrict our study to the most secure cases.

\subsection{$A_{\rm FUV}$ versus $\log(L_{\rm IR}/L_{\rm FUV})$}\label{sec:Auv-IRX}

\begin{figure}
\centering
\includegraphics[width=9cm]{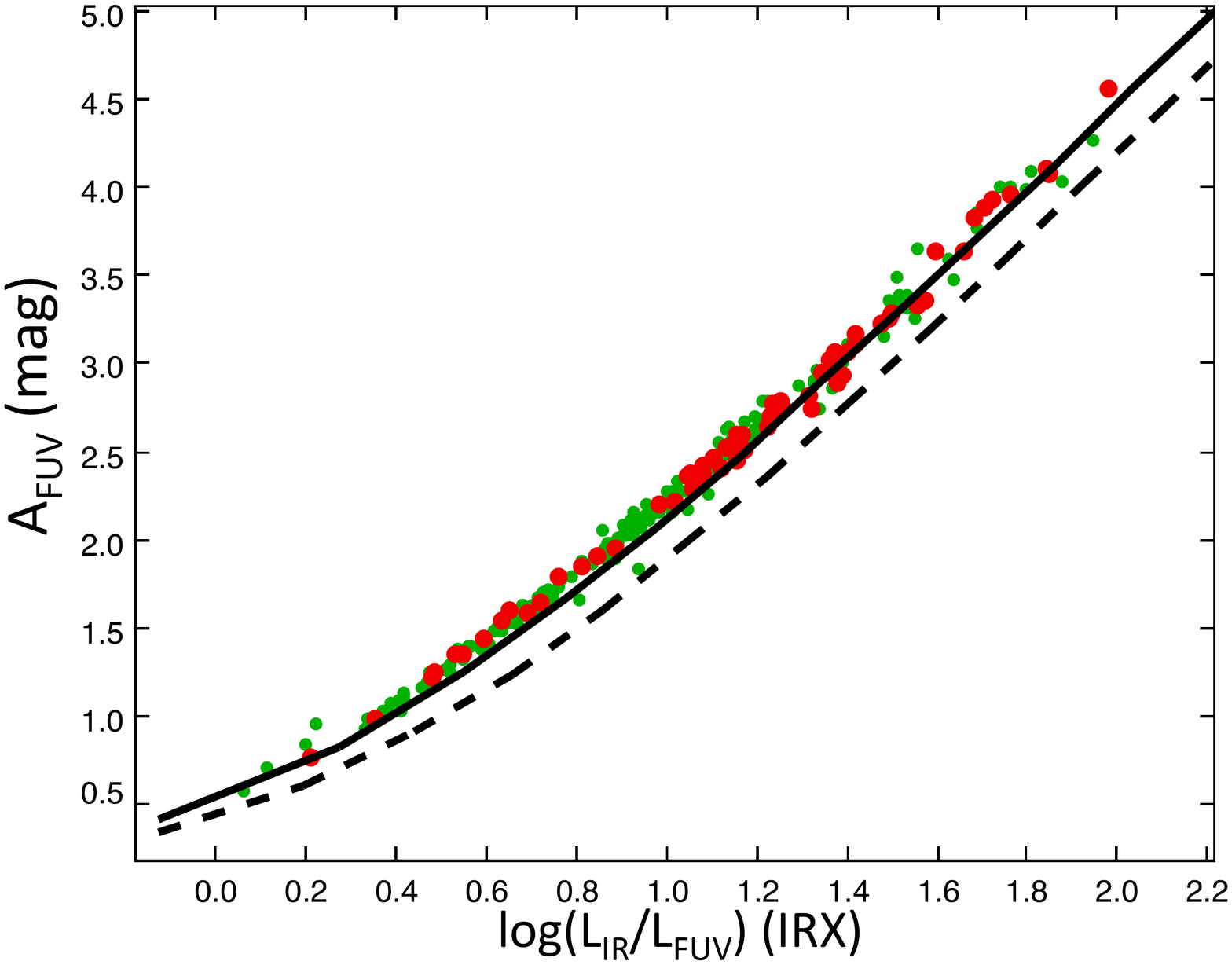}
\caption{Dust attenuation in FUV as a function of $\log(L_{\rm IR}/L_{\rm FUV})$~=~IRX. The symbols are the same as in Fig.~\ref{fig:mock}. The relation of \citet{overzier11} for local starbursts is plotted with a solid line, while the relation proposed by \citet{hao11} for local star-forming galaxies is plotted with a black dotted line.}
\label{fig:Auv-IRX}
\end{figure}

We adopt the formalism of \citet{hao11} (first proposed by \citet{Meurer99}) to relate $A_{\rm FUV}$ to $\log(L_{\rm IR}/L_{\rm FUV})$ (called IRX hereafter):
\begin{equation}\label{eq:Auv-IRX}
A_{\rm FUV} = 2.5 \log(1 + a_{\rm FUV} \times 10^{\rm IRX}),
\end{equation}
where $a_{\rm FUV} = 0.595$ for local starbursts \citep{overzier11} and 0.46 for normal star-forming galaxies \citep{hao11}. In Fig.~\ref{fig:Auv-IRX}, the results of our SED fitting are compared to Eq.~\ref{eq:Auv-IRX}. The local starburst relation is found to be consistent with our data, whereas the relation valid for nearby star-forming galaxies underestimates the dust attenuation in our high-redshift galaxies.

\subsection{A$_{\rm FUV}$-$\beta$ relation}\label{sec:Auv-beta}

\begin{figure}
\centering
\includegraphics[width=9cm]{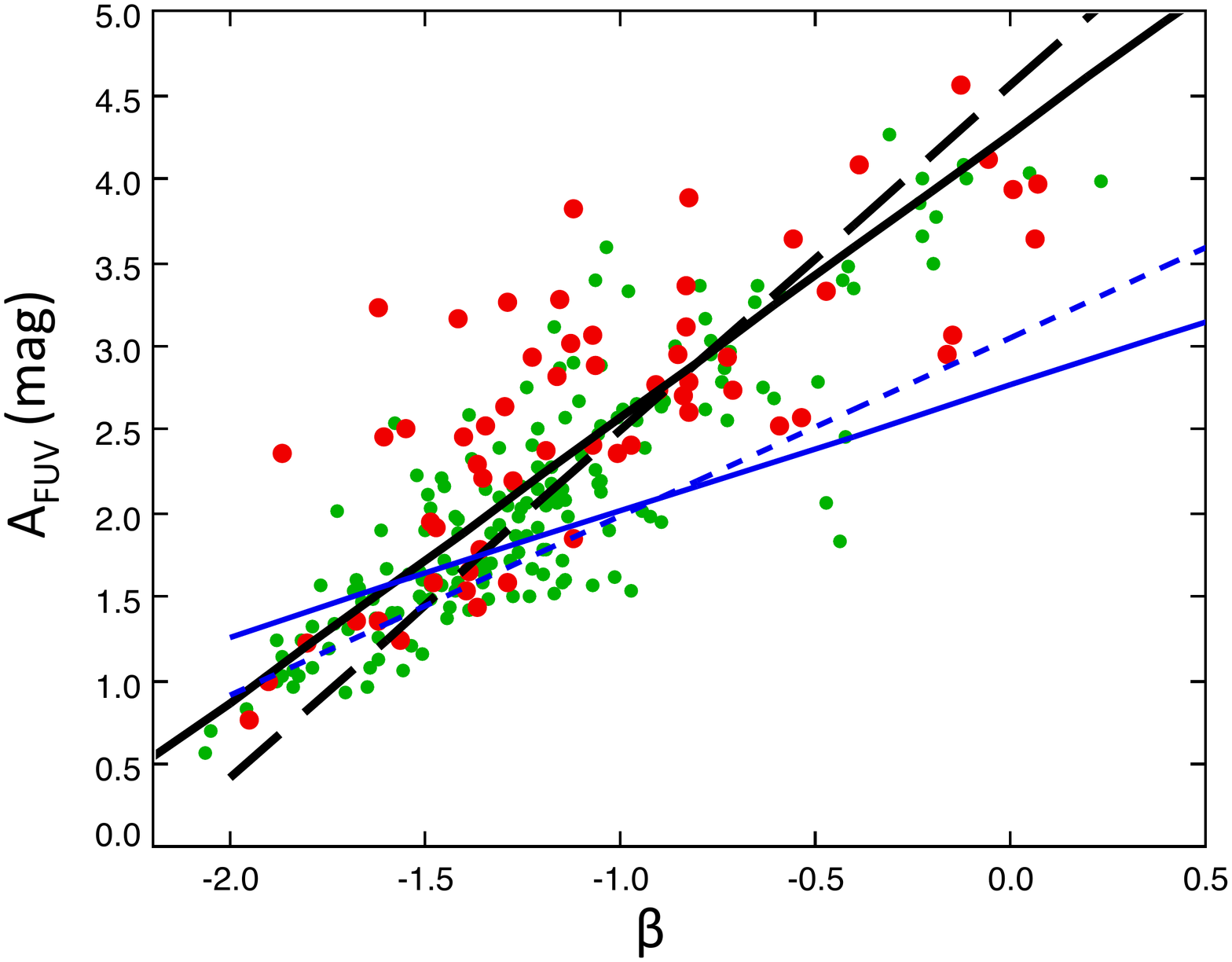}
\caption{Dust attenuation as a function of $\beta$ (same symbols as in Fig.~\ref{fig:mock}). The solid line corresponds to the relation found in this work: $A_{\rm FUV} = 1.70 \times (\beta + 2.5)$. The dashed line corresponds to the relation of \citet{overzier11} (their inner relation). The relation found by \citet{murphy11} for 24\,$\mu$m selected galaxies at $0.66 < z < 2.66$ and by \citet{buat11a} for nearby galaxies selected at 140\,$\mu$m are plotted as blue solid and dotted lines, respectively.}
\label{fig:Auv-beta}
\end{figure}

When the intrinsic UV spectrum is modelled by a power law of exponent $\beta$, the relation between the attenuated and unattenuated fluxes, $f(\lambda)$ and $f_0(\lambda)$ (expressed per unit wavelength), and the corresponding slopes $\beta$ and $\beta_0$ becomes
\begin{equation}\label{eq:beta-flux}
\beta-\beta_0 =  {{\log(f(\lambda_1)/f(\lambda_2))} \over {\log(\lambda_1/\lambda_2)}} - {{\log(f_0(\lambda_1)/f_0(\lambda_2))} \over {\log(\lambda_1/\lambda_2)}}.
\end{equation}
The wavelengths $\lambda_1$ and $\lambda_2$ are taken in the UV range, where the modelling by a power law is valid. Then, dust attenuation at $\lambda_1$ can be written:
\begin{equation}\label{eq:Auv-beta}
A(\lambda_1) =  {{2.5 \log(\lambda_2/\lambda_1)} \over {1-A(\lambda_2)/(A(\lambda_1)}} \times (\beta-\beta_0) = C(\lambda_1,\lambda_2) \times (\beta-\beta_0),
\end{equation}
where $A(\lambda_2)/A(\lambda_1)$ depends on the dust attenuation curve (Eq.~\ref{eq:attlaw}). In order to be consistent with our SED fitting results, we adopt $\lambda_1 = 1530\,\AA$ (corresponding to the FUV effective wavelength) and $\lambda_2 = 3000\,\AA$ to calculate $C(\lambda_1,\lambda_2)$. Using 1200 and 2500\,$\AA$ modifies the calculations by less than 1$\%$. We found that $\delta = -0.27 \pm 0.17$ for the entire sample which corresponds to $C(\lambda_1,\lambda_2) = 1.70$ (with a range of $1.46 - 1.97$ corresponding to the standard deviation for $\delta$). For comparison, $C(\lambda_1,\lambda_2) = 2.3$ for the pure Calzetti law ($\delta = 0$), 1.3 for the SMC and 1.8 for the LMC supershell using the extinction curves of \citet{gordon03}. We can estimate $\beta_0$ from Eq.~\ref{eq:Auv-beta}. Using $C(\lambda_1,\lambda_2) = 1.70$, we find $\langle\beta_0\rangle = -2.5 \pm 0.3$, while $\langle\beta_0\rangle = -2.7$ (resp. -2.3) for $C(\lambda_1,\lambda_2) = 1.46$ (resp. 1.97), consistent with the value expected for a young stellar population aged of $\sim 40$\,Myr (see Sect.~\ref{sec:SFH}). The corresponding relation between $\beta$ and $A_{\rm FUV}$ ($A_{\rm FUV} = 1.70 \times (\beta + 2.5)$) is plotted in Fig.~\ref{fig:Auv-beta} along with the data. We recall that $\beta$ is calculated on the entire UV range (Sect.~\ref{sec:sample}). The relation of \citet{overzier11} for local starburst galaxies (we consider their inner relation, very close to the original one obtained by \citet{Meurer99}) is also found to reproduce the general observed trend between $\beta$ and $A_{\rm FUV}$ with a formal scatter for both relations reaching 0.5\,dex. Their relation corresponds to $C(\lambda_1,\lambda_2) = 2.05$ and $\beta_0 = -2.19$ with the formalism adopted in this section. Flatter and very dispersed relations are found in the literature \citep[e.g.][and references therein]{buat11a,murphy11}. The relations found by \citet{buat11a} and \citet{murphy11} for IR-selected galaxies at low and high redshift are reported in Fig.~\ref{fig:Auv-beta}. Although a steeper attenuation curve may flatten the relation (cf. Eq.~\ref{eq:Auv-beta}), \citet{buat11a} show that a flat distribution with a large dispersion in the $A_{\rm FUV}$-$\beta$ diagram is consistent with an attenuation curve only slightly steeper than the Calzetti one and variations in the star formation history with a substantial contribution of old stars.

\subsection{IRX-$\beta$ relation}\label{sec:IRX-beta}

\begin{figure}
\centering
\includegraphics[width=9cm]{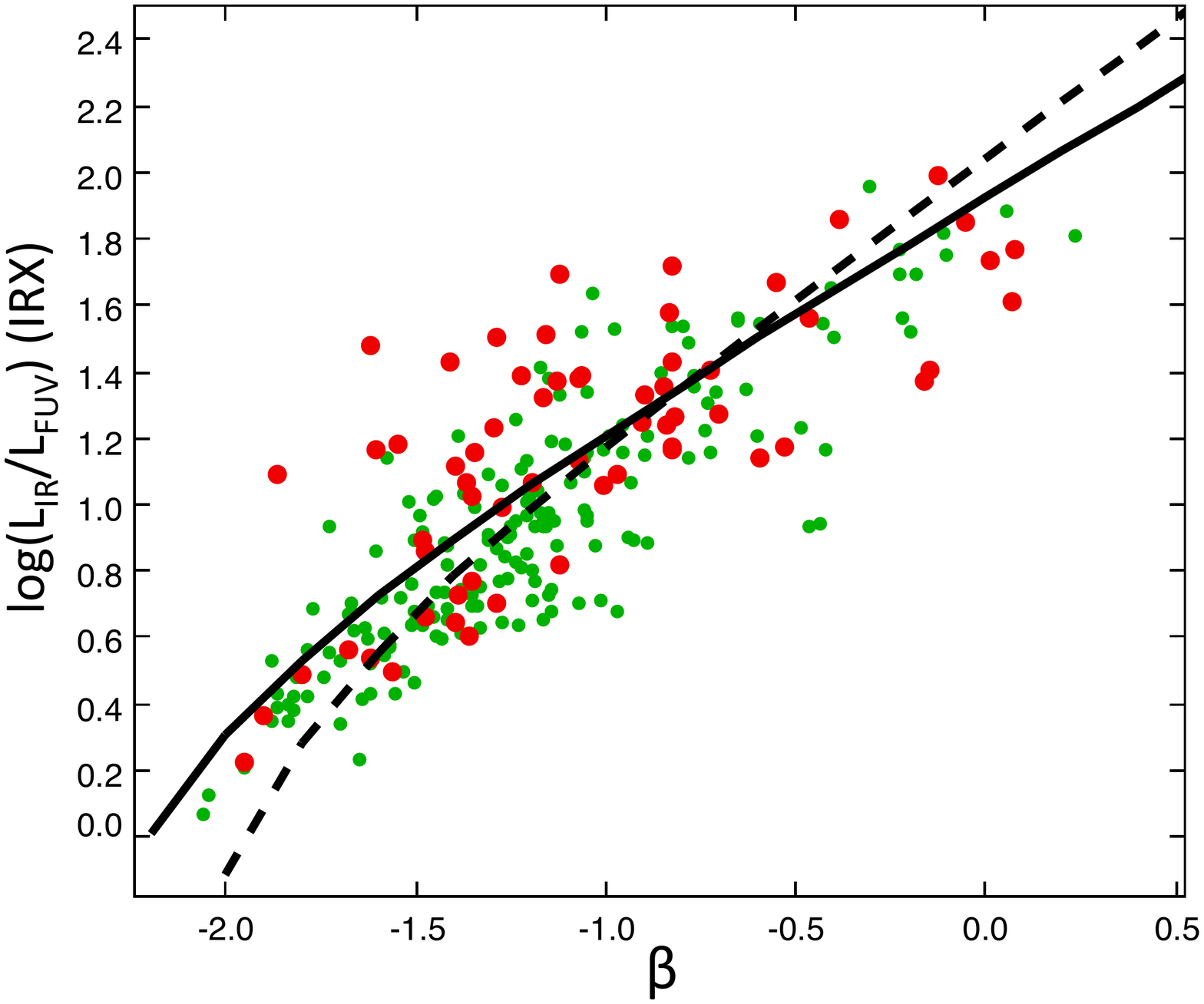}
\caption{The observed slope $\beta$ is plotted versus $\log(L_{\rm IR}/L_{\rm FUV}) = {\rm IRX}$. The symbols are the same as in Fig.~\ref{fig:mock}. The solid line is the relation found in this work, while the dashed line is that of \citet{overzier11}.}
\label{fig:IRX-beta}
\end{figure}

We combine Eq.~\ref{eq:Auv-IRX} for starbursts and Eq.~\ref{eq:Auv-beta} to deduce the IRX-$\beta$ relation:
\begin{equation}\label{eq:IRX-beta}
{\rm IRX} = \log(10^{0.4\times 1.7 (\beta+2.5)} - 1) - \log(0.595)
\end{equation}
The resulting curve is plotted with the data in Fig.~\ref{fig:IRX-beta}. The relation proposed by \citet{overzier11} for local starbursts is also shown. The modification of the slope of the attenuation curve (steeper than the Calzetti one and similar to that of the LMC supershell) and of the intrinsic UV slope (from $\sim -2.2$ to -2.5) modifies substantially the IRX-$\beta$ relation. The data are found to roughly follow both laws but a low value of $\beta_0$ is necessary to fit the small values of slope and attenuation.

\begin{figure}
\centering
\includegraphics[width=9cm]{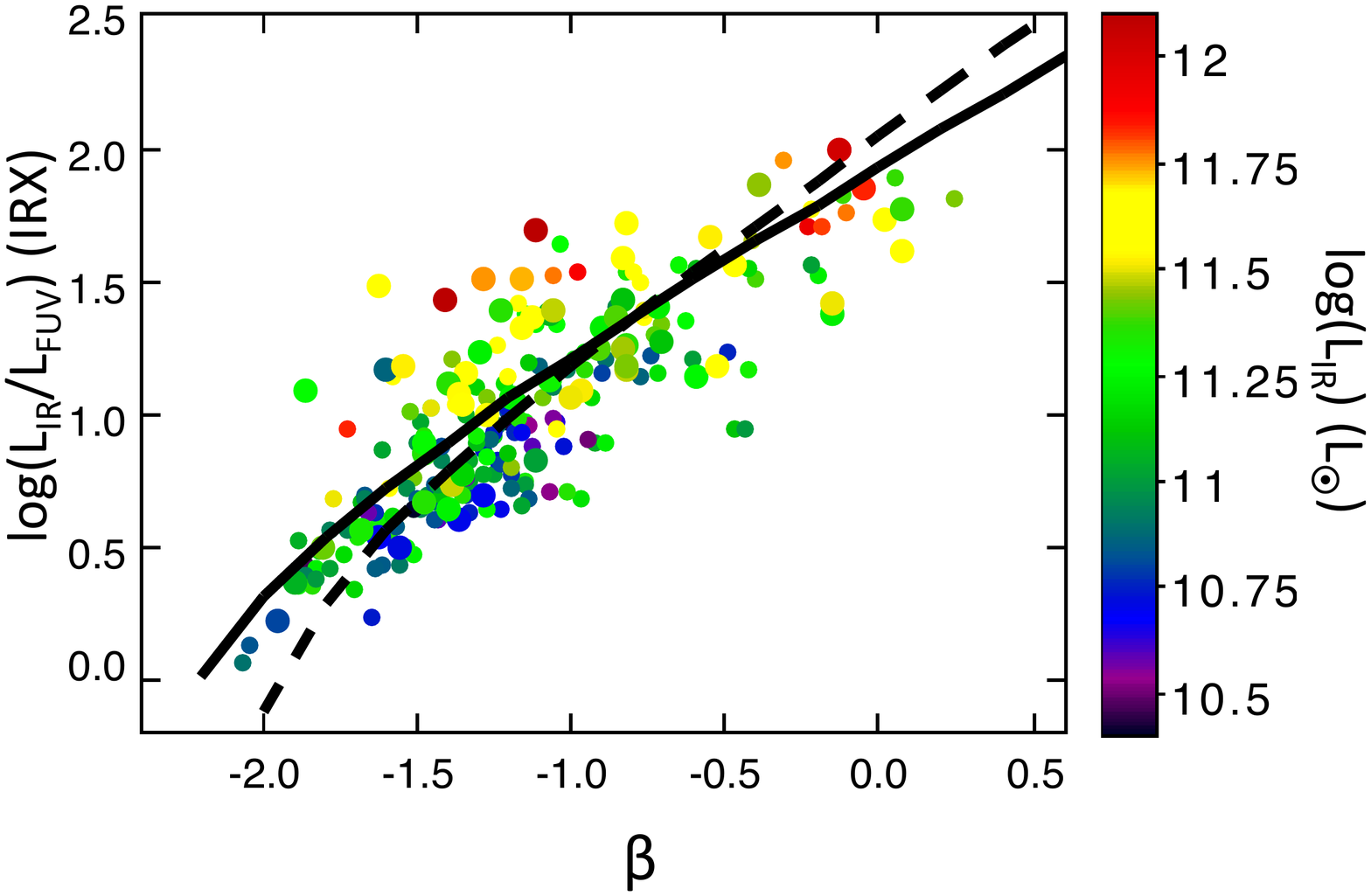}
\includegraphics[width=9cm]{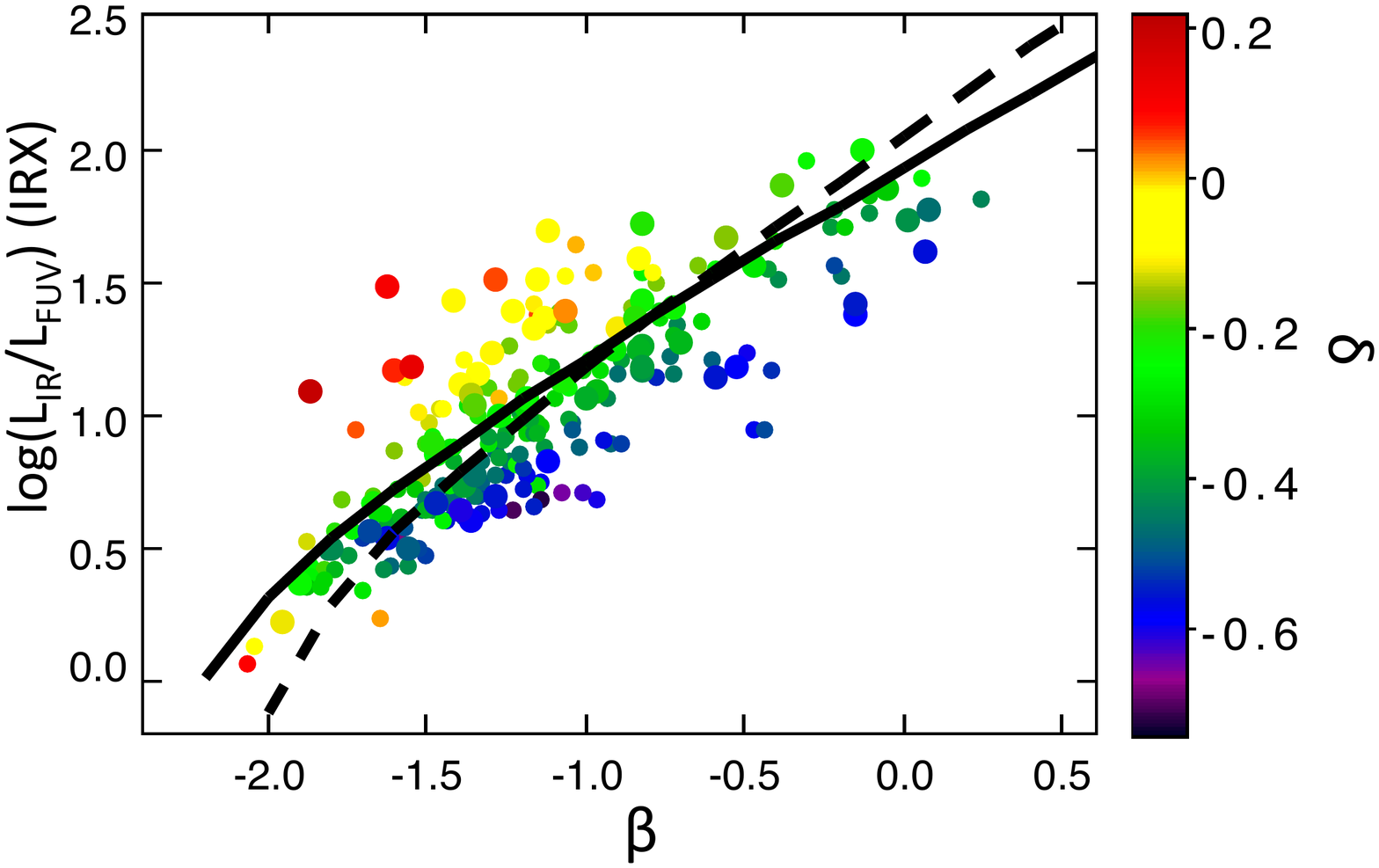}
\includegraphics[width=9cm]{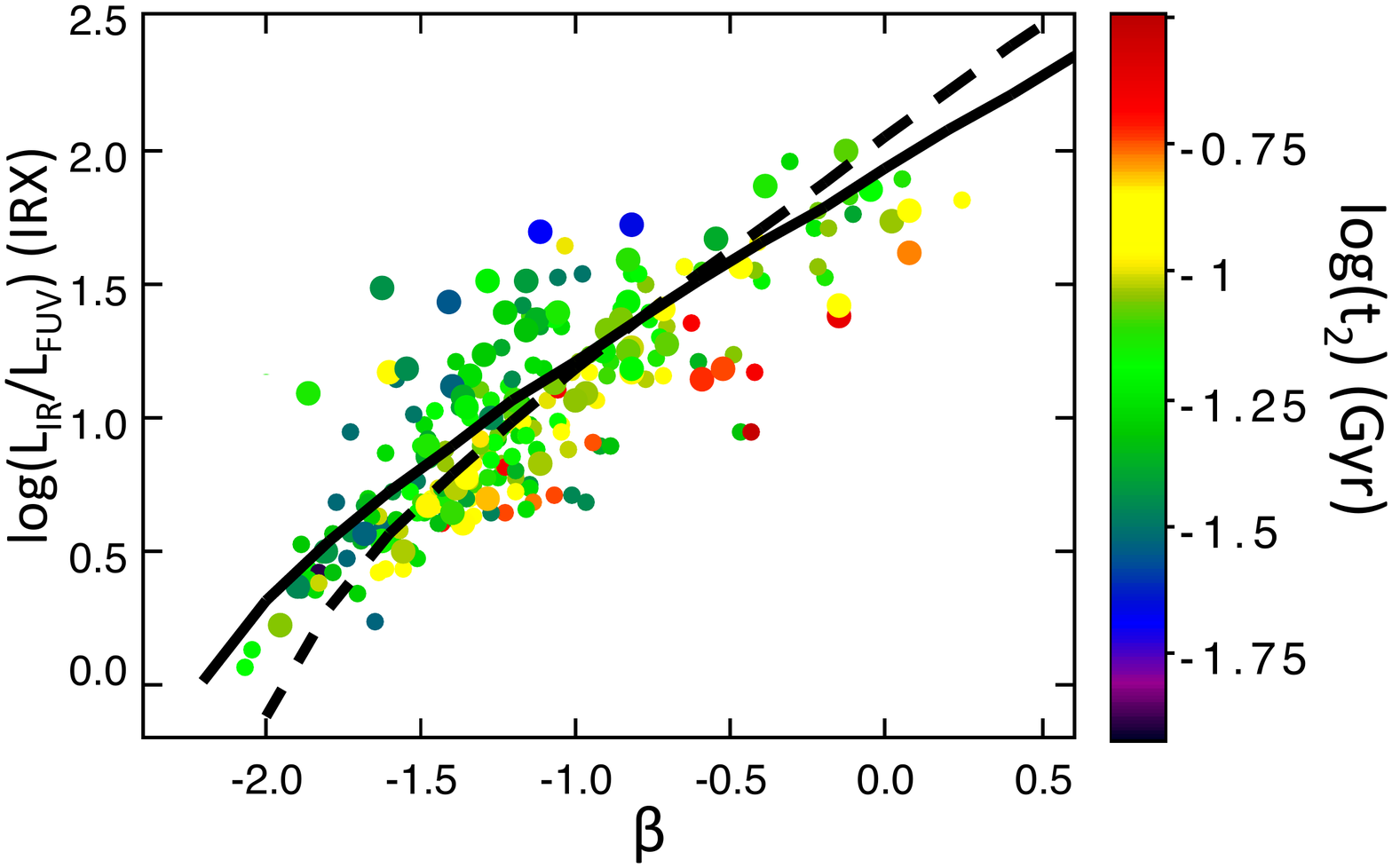}
\caption{The observed slope $\beta$ is plotted versus $\log(L_{\rm IR}/L_{\rm FUV}) = {\rm IRX}$. The dust luminosity $L_{\rm IR}$ (top panel), $\delta$ (median panel), and the age of the young stellar population $t_2$ (bottom panel) are colour coded. The solid line is the relation found in this work, while the dashed line is that of \citet{overzier11}.}
\label{fig:IRX-beta2}
\end{figure}

\begin{figure}
\centering
\includegraphics[width=9cm]{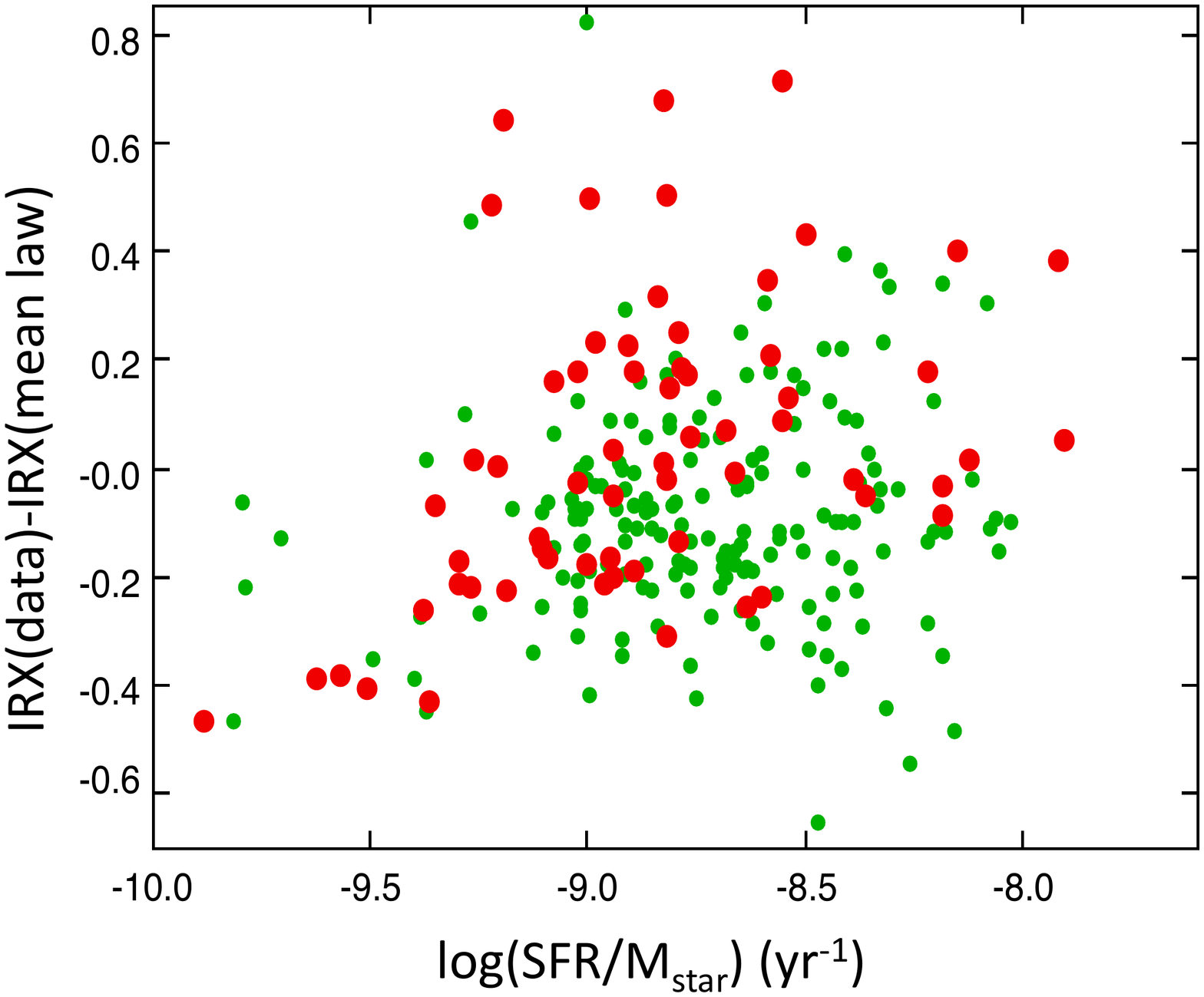}
\caption{Distance of the data points in the IRX-$\beta$ diagram from the mean law found in this work against the specific SFR. Symbols are the same as in Fig.~\ref{fig:mock}.}
\label{fig:delta-irx}
\end{figure}

The standard deviation from the Overzier et al. relation reaches 0.27\,dex and is reduced to 0.23\,dex when the relation derived in this work is used. It is similar to that found by \citet{Meurer99} for local starburst galaxies. At $z \approx 2$, \citet{reddy10} found that typical Lyman break galaxies roughly obey the local starburst IRX-$\beta$ relation with a formal scatter of 0.38\,dex, slightly higher than our value. This relatively low dispersion is probably due to our UV selection and the high star formation activity of our galaxies, since the original law was first proposed for local UV-selected starburst galaxies. Subpopulations of dusty, IR-luminous galaxies at low and high redshift are known to deviate from the original starburst relation with a higher IRX for a given slope $\beta$ \citep{goldader02,chapman05,howell10,papovich06}. \citet{penner12} select dust-obscured galaxies (DOGs) at $z \simeq 2$ and use GOODS-$Herschel$ data to study their IR properties. They find that DOGs lie on the left side of the IRX-$\beta$ relation, with a systematic increase of IRX with $\beta$.

The influence of various parameters on the dispersion found in Fig.~\ref{fig:IRX-beta} is explored in Fig~\ref{fig:IRX-beta2}. There is a global increase of IRX with $L_{\rm IR}$ for a given value of $\beta$. Such a trend was already reported at low and high redshift \citep[e.g.][and references therein]{howell10,chapman05}. We can also investigate the role of the variation of the attenuation curve ($\delta$) and of the star formation history ($\beta_0$). As shown in Fig.~\ref{fig:IRX-beta2}, the sources located above the mean law correspond to $\delta$ values less negative than our mean law ($\delta = -0.27$): they are fitted with a greyer attenuation curve. Such a trend with $\delta$ is easily explained: for a given slope $\beta$ these galaxies experience a larger dust attenuation than predicted by our mean law and the fit gives a less efficient reddening and thus a greyer attenuation. $\beta_0$ is not an output parameter of CIGALE but its influence can be investigated by using the age of the young stellar populations as a proxy: galaxies with the lowest age (and therefore a small value of $\beta_0$) are located well above the mean law and conversely the oldest populations (corresponding to large values of $\beta_0$) below it. As a consequence, the intrinsic value of $\beta_0$ may also play a role in the locus of the galaxies in the IRX-$\beta$ diagram (see also \citet{boquien12,heinis12}).

\citet{elbaz11} relate the starburst activity to a higher specific SFR and a high compactness of the star formation. In Fig.~\ref{fig:delta-irx}, the distance to the mean IRX-$\beta$ law is plotted versus the specific SFR of our galaxies. A weak but positive correlation is found (correlation coefficient $R = 0.20$), which is stronger ($R = 0.51$) when only galaxies detected with PACS are considered. This trend is consistent with a larger contribution of compact regions, which likely have greyer attenuation curves, with increasing specific SFR as proposed by \citet{elbaz11}.

\section{Conclusions}\label{sec:conclusions}

We have studied the dust attenuation properties in a sample of 751 galaxies with $0.95 < z <2.2$ observed with intermediate-band filters in order to sample correctly the UV rest-frame continuum. IR data from MIPS (290 galaxies) and PACS (76 galaxies) were used when available. Our results are:
\begin{enumerate}
\item SED fitting is performed with CIGALE. The dust attenuation law is described with two free parameters: its steepness ($\delta$) and the amplitude of a bump at 2175\,$\AA$ ($E_{\rm b}$). The availability of IR data improves the determination of all the parameters related to dust attenuation. Global parameters such as the SFR, stellar mass, or the amount of dust attenuation are well determined, whereas the detailed star formation history is ill-constrained. Dust attenuations, total IR luminosities, and $\delta$ are found to be slightly overestimated in the regime of low attenuation and luminosity when IR data are missing.
\item The dust attenuation in the FUV is found to have a large  dispersion for low $L_{\rm FUV}$ ($L_{\rm FUV} < 10^{10}\,L_{\odot}$). The mean value of $A_{\rm FUV}$ decreases from 2.4 to 1.5\,mag when $L_{\rm FUV}$ increases from $10^{9.7}$ to $10^{10.8}\,L_{\odot}$. The dust attenuation is found to increase with stellar mass.
\item The parameters describing the dust attenuation curve, $E_{\rm b}$ and $\delta$, are found to span a large range of values. This is likely to reflect intrinsic variations of the attenuation curve among galaxies and for different environments within a galaxy. The presence of a bump is confirmed in 20$\%$ of the sample, at $z < 1.5$ for 90$\%$ of the confirmed detections. A dust attenuation law steeper than that of \citet{calzetti00} ($\delta < 0$) is also confirmed for 20$\%$ of the sample. These percentages increase to 40$\%$ for galaxies with IR detections. The mean values of $E_{\rm b}$ and $\delta$ derived for the whole sample are similar to those found for the extinction curve of the LMC supershell. Finally, $E_{\rm b}$ is found to be anti-correlated with the specific SFR of galaxies.
\item The relations between $A_{\rm FUV}$ and $\beta$ (the slope of the UV continuum) and IRX ($\log(L_{\rm IR}/L_{\rm FUV})$) and $\beta$ are derived for our mean attenuation curve and for galaxies of our sample detected in IR. The intrinsic slope of the UV continuum, $\beta_0$, is found equal to -2.5 and the galaxies follow the IRX-$\beta$ relation with a low dispersion. Galaxies found above the average IRX-$\beta$ relation have a high IR luminosity and a dust attenuation curve slightly greyer than our mean law, which can be due to a higher compactness of the star formation regions. The age of the young stellar population (i.e. the intrinsic shape of the UV continuum) is likely to also play a role: for a given $\beta$, IRX increases when the age of the young stellar population decreases.
\end{enumerate}

\begin{acknowledgements}
This work is partially supported by the French National Agency for research (ANR-09-BLAN-0224). Support for the US-based co-authors was provided by NASA through an award issued by JPL/Caltech.
PACS has been developed by a consortium of institutes led by MPE (Germany)
and including UVIE (Austria); KU Leuven, CSL, IMEC (Belgium); CEA,
LAM (France); MPIA (Germany); INAFIFSI/ OAA/OAP/OAT, LENS, SISSA
(Italy); IAC (Spain).
This development has been supported by the funding agencies
BMVIT (Austria), ESA-PRODEX (Belgium), CEA/CNES (France), DLR
(Germany), ASI/INAF (Italy), and CICYT/MCYT (Spain).
\end{acknowledgements}

\end{document}